\documentclass[a4paper]{iopart}
\pdfoutput=1
\usepackage[utf8]{inputenc}
\usepackage{amsfonts,amsmath,amssymb,graphicx}
\usepackage{fancyhdr}
\usepackage{xcolor}
\usepackage[square,numbers,sort&compress]{natbib}

\usepackage[colorlinks]{hyperref}
\definecolor{darkred}{rgb}{0.5,0,0}
\definecolor{darkgreen}{rgb}{0,0.5,0}
\definecolor{darkblue}{rgb}{0,0,0.5}
\hypersetup{colorlinks,
linkcolor=darkred,
filecolor=darkgreen,
urlcolor=darkblue,
citecolor=darkgreen}

\usepackage{orcidlink}
\usepackage{bbold}
\usepackage{braket}
\usepackage{amsmath}
\usepackage{mathtools}
\usepackage{braket}
\usepackage{amssymb}
\usepackage{overpic}
\usepackage{float}
\usepackage{color}
\usepackage{braket}
\usepackage[english]{babel}
\usepackage{cleveref}

\newcommand{\transpose}{{\scriptscriptstyle \mathrm{T}}}

\newcommand{\dd}{\text{d}}
\newcommand{\eff}{\text{eff}}

\newcommand{\Ham}{H}

\newcommand{\dt}{\text{d}t}

\begin{document}
\title{Long-lasting Topological Entanglement in a Monitored Rashba Nanowire}

\author{Emanuele Guida~\orcidlink{0009-0005-5742-8360}}
\address{Dipartimento di Fisica ``E. Pancini'', Università di Napoli Federico II, I-80126 Napoli, Italy}
\author{Giulia Salatino~\orcidlink{0009-0005-6913-6920}}
\ead{giulia.salatino@unina.it}
\address{Scuola Superiore Meridionale, Via Mezzocannone, 4 80138, Napoli, Italy}
\author{Gianluca Passarelli~\orcidlink{0000-0002-3292-0034}}
\address{Dipartimento di Fisica ``E. Pancini'', Università di Napoli Federico II, I-80126 Napoli, Italy}
\author{Angelo Russomanno~\orcidlink{0009-0000-1923-370X}}
\address{Dipartimento di Fisica ``E. Pancini'', Università di Napoli Federico II, I-80126 Napoli, Italy}
\author{Procolo Lucignano~\orcidlink{0000-0003-2784-8485}}
\address{Dipartimento di Fisica ``E. Pancini'', Università di Napoli Federico II, I-80126 Napoli, Italy}
\begin{abstract}
We study the topological properties of a monitored Rashba chain along quantum-jump trajectories, investigating the persistence of the initial topological value of the disconnected entanglement entropy (DEE). We find that the DEE persists in its topological value for a time linear in the system size, even if the dissipation acts on the boundary and affects the topological Majorana modes. The reason for this phenomenon lies in the absence of particle conservation and in the degeneracy of the topological manifold, allowing the monitoring to let the system switch between different topological states -- alternatively creating and annihilating a Majorana mode -- while producing a poisoning of finite-energy ballistically propagating quasiparticles that eventually destroy the topological entanglement structure.
\end{abstract}

\section{Introduction}

Topological phases are one of the most important developments in condensed matter physics in recent years~\cite{RevModPhys.89.040502}. They cannot be characterized by local order parameters, but are instead encoded in global or nonlocal properties. Celebrated examples are the Su-Schrieffer-Heeger (SSH)~\cite{ssh_1979,ssh_1980,Lieu_2018} and Kitaev~\cite{Kitaev_2001, Alicea_2012,10.21468/SciPostPhysLectNotes.82} models, which host boundary topological modes.

In this context, the disconnected entanglement entropy (DEE)~\cite{Dalmonte_PhysRevB.101.085136, DEE_ssh} has been proposed as a probe of symmetry-protected topological phases. It is defined as a combination of entanglement entropies of different subsystems, and it is nonvanishing only in the presence of  non-local topological boundary modes. It has been shown to detect not only non-local entanglement in symmetry-protected topological phases~\cite{torre2023longrangeentanglementtopologicalexcitations}, but also Majorana zero modes in semiconductor-superconductor heterostructures~\cite{arora2023conclusivedetectionmajoranazero}. Since it is built from entanglement entropies, the DEE can also be experimentally accessible~\cite{Zoller_PRX_2016, Zoller_PRL_2018, Zoller_Science_2019}. The DEE has also been studied in quenched dynamics: consistently with the Lieb-Robinson bound~\cite{cmp/1103858407},  correlations spread ballistically, and a time linear in the system size is required before the two boundaries become dynamically connected, and the delocalized topological superposition is destroyed~\cite{Dalmonte_PhysRevB.101.085136, Micallo_2020, Mondal_2022}. 

In recent years, the study of topological phases has been extended beyond isolated Hermitian systems to non-Hermitian Hamiltonians~\cite{Ashida_2020,okuma2023nonhermitian}, open quantum systems governed by Lindblad dynamics~\cite{Diehl_2011,bardyn2013topologybydissipation,cooper,Altland_2021} and monitored quantum systems~\cite{Viotti2023geometricphases, romito2023topologymonitored, articolo_Giulia}. Non-unitary evolution can give rise to phenomena without a direct counterpart in closed systems, including the breakdown of conventional bulk-boundary correspondence and the non-Hermitian skin effect, which can be understood within the framework of spectral topology~\cite{Ashida_2020}. In open systems, topology may characterize different objects, such as the spectrum of an effective non-Hermitian Hamiltonian or of a Liouvillian, the stationary state, or the pure states generated along individual quantum trajectories. These notions need not be equivalent. This has motivated the introduction of several complementary probes of topology, including non-Hermitian topological invariants~\cite{Kunst_2018,Gong_2018,Shen_2018}, the Uhlmann phase~\cite{Delgado_2014,Viyuela_2018,carollo2017uhlmanncurvaturedissipativephase}, the ensemble geometric phase~\cite{Bardyn_2018,Unayan_2020}, mixed-state topological order parameters~\cite{huang2024mixedstatetopologicalorder}, long-time entanglement negativity~\cite{chen2024universalentanglementrevivaltopological}, and transport observables such as imbalance and current anomalies~\cite{nava2023,campagnano2024}.

In what follows, we focus on the fate of symmetry-protected topological phases under continuous monitoring. Monitored dynamics gives access to the stochastic evolution of pure states along individual quantum trajectories. This makes it possible to study nonlinear quantities, such as entanglement entropies, whose trajectory averages cannot in general be reconstructed from the ensemble-averaged density matrix. See, for instance, the reviews~\cite{Plenio_1998,Daley_2014,SciPostPhysLectNotes.99}.

In our previous work on the monitored SSH chain~\cite{articolo_Giulia}, we studied the DEE under different quantum-jump dynamics. We found that the transient stability of the entangled edge modes is controlled primarily by the spatial profile of the dissipation: when the boundaries are protected from quantum jumps, the DEE remains close to its quantized topological value for a time that increases linearly with the system size, whereas jumps acting directly on the boundaries rapidly destroy the edge entanglement. The analysis of individual trajectories further showed how this loss can be traced to specific local jump events.

The present work extends this trajectory-resolved approach from the SSH chain to a superconducting Rashba nanowire hosting Majorana boundary modes~\cite{Alicea_2012,majorana_quasiparticles_in_condensed_matter_2017,Oreg_2010,Leijnse_2012,PhysRevLett.105.077001,doi:10.1126/science.1222360,Das2012}. We consider a Rashba nanowire coupled to a dissipative environment described by a Lindbladian inspired by Ref.~\cite{ar_nanowire}. We describe the effect of the environement as an external monitoring and unravel the Lindblad dynamics into a quantum-jump stochastic trajectory scheme~\cite{Plenio_1998,Daley_2014}. This allows us to follow the DEE along pure-state evolution of individual trajectories. 

Unlike the SSH model, the superconducting nanowire does not conserve particle number. Its topological phase is characterized by a pair of ground states with different fermion parity, as in the Kitaev chain, which provides the low-energy description of the model~\cite{Alicea_2012, hassler2014majoranaqubits}. This difference is central for the monitored dynamics. It allows us to investigate whether the non-local entanglement structure associated with a Majorana boundary modes can still be detected when local quantum jumps generate quasiparticle excitations, a process related to quasiparticle poisoning, which is a central decoherence mechanism in Majorana devices~\cite{RainisLoss2012,Karzig2021,Albrecht2017}.

Our main result is that, in sharp contrast to the SSH model, the topological value of the DEE survives for a time scaling linearly with system size, independently of whether dissipation acts in the bulk, at the boundaries, or everywhere. This behavior can be understood through a minimal Kitaev-chain picture~\cite{hassler2014majoranaqubits}. Along a quantum-trajectory, boundary jumps change the fermion parity, alternatively creating and annihilating the zero-energy topological modes, thus moving the system between the two states of the topological ground-state manifold. At the same time, the same boundary jumps generate finite-energy contributions localized near the edge. The associated entanglement spreads through the chain and reaches the regions entering the DEE only after a time proportional to the system size. Physically, this time is the scale at which the propagating quasiparticles let the boundaries start cross talking with each other, disrupting topological entanglement.

Thus, unlike in the SSH chain, boundary dissipation does not simply erase an occupied edge mode. Owing to ground-state degeneracy and particle-number nonconservation, it first induces parity switches within the topological manifold. The eventual loss of the DEE is instead governed by the spreading of the entanglement generated by the finite-energy contributions produced at the edge. In this sense, the DEE provides a direct probe of how local poisoning events are converted into the delayed decay of nonlocal topological entanglement,with a lifetime that grows linearly with system size and therefore diverges in the thermodynamic limit.

The paper is organized as follows. In Section~\ref{sec2} we introduce the Hamiltonian of the superconducting Rashba nanowire, and its topological properties. In Section~\ref{sec3} we introduce the quantum-trajectory formalism, focusing on the quantum-jump unraveling. We then present the DEE and the method to compute its time evolution. 
In Section~\ref{sec5}, we show that the topological value of the DEE persists for a time linear in the system size, independently of where the dissipation acts. In Section~\ref{sec:kitaev_benchmark}, we interpret this behavior through a simple analytical picture based on the low-energy Kitaev-chain description. In Section~\ref{sec:conclusions}, we draw our conclusions.

\section{The model}
\label{sec2}
We consider a Rashba nanowire with open boundary conditions, consisting of $N$ lattice sites, in proximity to a conventional spin-singlet $s$-wave superconductor and subject to a magnetic field applied along the wire~\cite{Alicea_2012,majorana_quasiparticles_in_condensed_matter_2017,Oreg_2010,Leijnse_2012,PhysRevLett.105.077001,doi:10.1126/science.1222360,Das2012}, which has been also generalized to  $d$-wave superconductors \cite{Lucignano2012YBCO,Lucignano2013SQUID}.  The second-quantized mean-field Hamiltonian reads
\begin{equation}
    \begin{split}
        H=&-\mu\sum_{i=1,\alpha}^Nc_{i\alpha}^\dagger c_{i\alpha}
        +t_h\sum_{i=1,\alpha}^{N-1}\left(c_{i\alpha}^\dagger c_{i+1\alpha}+\text{h.c.}\right)
        -i\lambda_R\sum_{i=1,\alpha,\beta}^{N-1}\left(c_{i\alpha}^\dagger (\sigma_z)_{\alpha\beta}c_{i+1\beta}+\text{h.c.}\right)\\
        &+B_x\sum_{i=1,\alpha,\beta}^{N}c_{i\alpha}^\dagger (\sigma_x)_{\alpha\beta}c_{i\beta}
        +\Delta\sum_{i=1}^N\left(c_{i\uparrow}^\dagger c_{i\downarrow}^\dagger+\text{h.c.}\right)\;,
    \end{split}
    \label{Nanowire_rasba_Bx}
\end{equation}
assuming spin indices $\alpha,\beta\in\{\uparrow,\downarrow\}$. Here $\mu$, $t_h$, $\lambda_R$, $B_x$, and $\Delta$ denote the chemical potential, hopping amplitude, Rashba spin-orbit coupling strength, magnetic-field amplitude, and proximity-induced $s$-wave pairing, respectively.  Moreover, $c_{i\alpha}$ ($c_{i\alpha}^\dagger$) annihilates (creates) an electron on site $i$ with spin $\alpha\in\{\uparrow,\downarrow\}$, while the Pauli matrices $\boldsymbol{\sigma}$ act on the spin degrees of freedom.

This model Hamiltonian exhibits Kitaev-like physics -- in the low-energy regime it maps onto an effective 1D spinless $p$-wave superconductor~\cite{Alicea_2012,Oreg_2010}. When prepared in its topological phase in open boundary conditions, it supports two exponentially localized zero-energy edge states, namely Majorana zero modes~\cite{Kitaev_2001}. The bulk gap closes at the critical fields
\begin{equation}
    B_{\pm}=\sqrt{(\mu\pm 2t_h)^2+\Delta^2}
    \;,
    \label{campi_critici}
\end{equation}
and the system is in the topological phase for
\begin{equation}
    B_{-}<B_x<B_{+}\;,
\end{equation}
where it hosts two Majorana zero modes~\cite{Oreg_2010,PhysRevB.107.035427}. In the following, we set $t_h=\Delta=\mu=1$ and $\lambda_R=0.5$. 

We introduce the coupling to the environment by treating the system as an open Markovian quantum system whose dynamics is governed by the Lindblad equation~\cite{breuer,lindblad1975generators}
\begin{equation}
    \dot{\rho}= -i\left[H,\rho\right]+\sum_\mu  \left(L_{\mu}\rho L_{\mu}^\dagger-\frac{1}{2}\left\{L_\mu^\dagger L_\mu, \rho\right\}\right)\;.
    \label{lindHeff}
\end{equation}
The operators $\{L_\mu\}$ are the Lindblad, or jump, operators. We choose local spin-dependent particle-loss processes,
\begin{equation}
    L_{i\uparrow}=\sqrt{\gamma_\uparrow}c_{i\uparrow}\;, \hspace{5mm} 
    L_{i\downarrow}=\sqrt{\gamma_\downarrow}c_{i\downarrow}\;, \hspace{5mm} i=1,\dots,N\;.
    \label{eq:jump_operator}
\end{equation}
We will study different monitoring settings depending on the sites \(i\) where jump operators act, including homogeneous dissipation, bulk-only and edge-only configurations in which the index \(i\) will span a shorter range of sites localized in the bulk or at the edges, respectively.
These jump operators locally break fermionic parity and therefore allow the generation of single quasiparticles, providing thereby a natural effective description of quasiparticle poisoning processes. In the case of spatially uniform dissipation, this choice preserves translation invariance while retaining the spin dependence of the losses. Since the spin orbit coupling and the magnetic field break spin reversal symmetry, one may expect that also monitoring acts differently in the two spin channels, hence we choose $\gamma_\uparrow\neq \gamma_\downarrow$, and throughout this work we set $\gamma_\uparrow=1$ and $\gamma_\downarrow=0.9$, having checked numerically that moderate variations of these parameters do not qualitatively affect the observed phenomenology.

In what follows, we prepare the initial state deep in the topological phase and study how its non-local properties evolve under the dissipative dynamics. For this reason, unless otherwise stated, we fix $B_x=2$, which lies within the critical boundaries defined in  Eq. \ref{campi_critici}, for our choice of parameters.  
In addition to dissipation, we include  the effects of static disorder via a site-dependent chemical potential
\begin{equation}
    \mu_i=\mu+\delta\mu_i\;,
\end{equation}
where $\delta\mu_i$ is drawn from a uniform distribution in the interval $[-W,W]$. Since our focus is on the weak-disorder regime, we restrict the disorder strength to the range $W \in [0,1]$. For the value of $\mu$ considered here, this choice ensures that the local chemical potential $\mu_i$ remains positive at all sites.

\section{Methods}
\label{sec3}

We now introduce the tools used to characterize the monitored dynamics considered below. In particular, we formulate the dynamics in terms of quantum-jump unraveling of the Lindblad master equation.

\subsection{Quantum-jump unraveling}

The Lindblad dynamics can be unraveled into quantum trajectories, i.e., stochastic pure-state evolutions whose ensemble average reproduces the time-evolution of density matrix~\cite{carmichael2013statistical,Daley_2014,Plenio_1998,Molmer:93,breuer}. In continuously monitored systems, each trajectory is associated with a measurement record and thus has a direct experimental interpretation~\cite{lesanovsky2022fcs,potts2024fcs}.

In the quantum-jump unraveling, a trajectory consists of a smooth non-unitary evolution generated by an effective non-Hermitian Hamiltonian, interrupted by stochastic jumps associated with the Lindblad jump operators in Eq.~\eqref{eq:jump_operator}~\cite{Jacobs_2014,Yip_2018,wiseman2010quantum}. This description is equivalent to the density-matrix one for linear observables, but noy for nonlinear entanglement measures. In this context, the DEE is evaluated on pure states along individual trajectories and then averaged over the ensemble; applying it directly to the mixed density matrix would instead mix quantum and classical correlations.

The monitored dynamics is governed by the stochastic Schr{\"o}dinger equation
\begin{equation}
   \dd\ket{\psi(t)}=
   \left(-i H_{\mathrm{eff}}+\frac{1}{2}\sum_k \braket{L_k^\dagger L_k}_t\right)\dd t\,\ket{\psi(t)}
   +\sum_k \dd N_k(t)\left(\frac{L_k}{\sqrt{\braket{L_k^\dagger L_k}_t}}-\mathbf{1}\right)\ket{\psi(t)}\;,
   \label{equivalenza}
\end{equation}
where \(\ket{\psi(t)}\) denotes the system wavefunction (trajectory) and
\[
\braket{L_k^\dagger L_k}_t=\bra{\psi(t)}L_k^\dagger L_k\ket{\psi(t)}\;.
\]
The effective non-Hermitian Hamiltonian is
\begin{equation}
    H_{\mathrm{eff}}=H-\frac{i}{2}\sum_k L_k^\dagger L_k\;,
    \label{efficace}
\end{equation}
to which the c-number term
\[
\frac{i}{2}\sum_k \langle L_k^\dagger L_k \rangle_t
\]
is added in Eq.~\eqref{equivalenza}, ensuring norm preservation between jumps. The stochastic increments \(\dd N_k(t)\) are Poisson variables satisfying
\begin{equation}
    \dd N_k(t)=
    \left\{
    \begin{array}{ll}
    1 & \text{with probability } \braket{L_k^\dagger L_k}_t\,\dd t\\
    0 & \text{with probability } 1-\braket{L_k^\dagger L_k}_t\,\dd t
    \end{array}
    \right.\;,
\end{equation}
so that each jump corresponds to the action of \(L_k\) on the conditioned state.

Since the initial state is Gaussian, and the dynamics is generated by a quadratic Hamiltonian together with jump operators linear in the fermionic fields, the state remains Gaussian along each quantum trajectory~\cite{bettman2025gaussiantraj}. Therefore, each trajectory can be equivalently characterized by its correlation matrix, which can be evolved directly in time. This free-fermion formulation provides an efficient numerical approach, since the number of dynamical variables grows only polynomially with the system size, rather than exponentially as in the full Hilbert-space description. This, in turn, allows us to access large systems and approach the thermodynamic limit, where non-local entanglement-based order parameters become sharply defined. Details on the evolution of the correlation matrix along quantum-jump trajectories are reported in~\ref{appB}. 

\subsection{Disconnected Entanglement Entropy}\label{dee:sec}

Tracking the entanglement dynamics of the monitored system provides a complementary route to characterize its dynamical topological properties. To this end, we employ the DEE, \(S^D\), introduced in Refs.~\cite{Micallo_2020,Dalmonte_PhysRevB.101.085136}. The DEE is the one-dimensional counterpart of disconnected-entropy constructions originally developed to isolate universal entanglement contributions in topologically-ordered phases~\cite{preskill2006tee,wen2006tee, zeng2018quantum}, and it is designed to detect symmetry-protected topological phases, where the relevant non-local contribution is associated with boundary modes under open boundary conditions. In particular, for one-dimensional gapped systems in a symmetry-protected topological phase, it was shown in Refs.~\cite{Micallo_2020,Dalmonte_PhysRevB.101.085136} that
\begin{equation}
    \lim_{N\to\infty} S^D = \mathcal{D}\log 2\;,
\end{equation}
where \(\mathcal{D}\) is the number of Majorana zero modes selected by the bulk-edge correspondence, while it is zero in the trivial phase. Hence, \(S^D\) is a robust non-local order parameter in the thermodynamic limit. 

\begin{figure}
    \centering
    \includegraphics[width=0.9\linewidth]{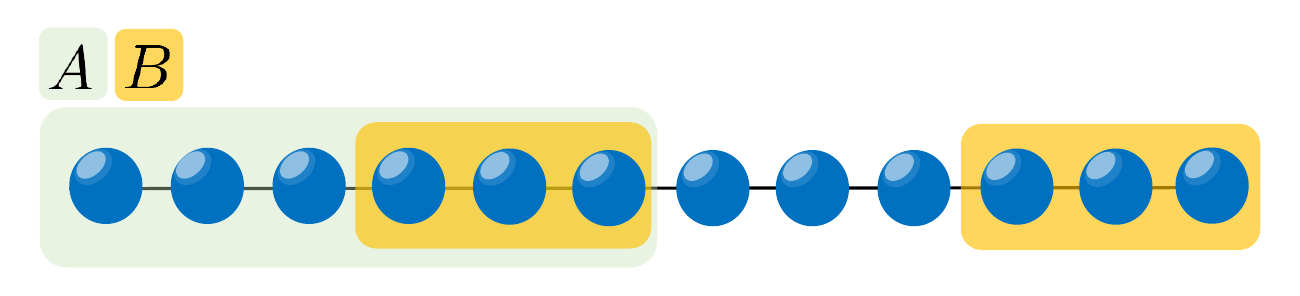}
    \caption{Sketch of the partition used to compute the disconnected entanglement entropy. 
The green shaded region denotes subsystem \(A\), while the yellow shaded regions denote subsystem \(B\). 
Blue spheres represent Dirac fermions localized on the sites of the chain. }
    \label{fig:ssh_sketch}
\end{figure}

\begin{figure}[ht]
    \centering
    \includegraphics[width=0.8\linewidth]{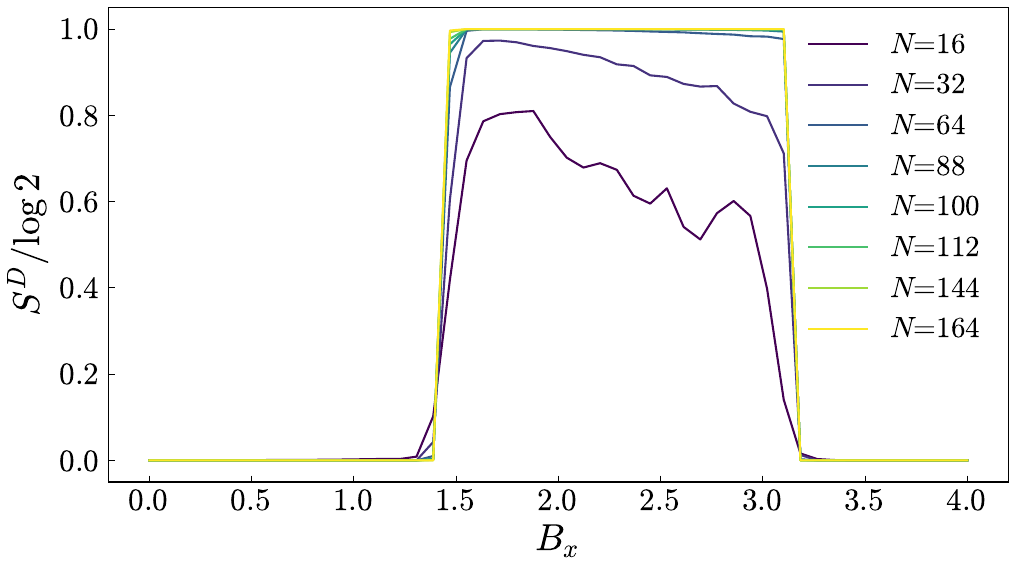}
    \caption{DEE of the ground state of the isolated Rashba nanowire Hamiltonian as a function of the external magnetic field, for growing system sizes.}
    \label{fig:dee_gs}
\end{figure}

The key idea is to combine entanglement entropies of regions with different connectivity so as to cancel non-universal contributions, such as the leading area-law term, and isolate the non-local entanglement encoded by the edge degrees of freedom. For a subsystem \(X\), the von Neumann entanglement entropy is
\begin{equation}
    S(X)=-\Tr\!\left(\rho_X \log \rho_X\right)\;,
\end{equation}
where \(\rho_X=\Tr_{\overline{X}}\rho\) is the reduced density matrix of \(X\). The DEE is then defined as
\begin{equation}
    S^D=S(A)+S(B)-S(A\cup B)-S(A\cap B)\;,
    \label{DEE}
\end{equation}
where \(A\) and \(B\) are the disconnected regions shown in Fig.~\ref{fig:ssh_sketch}.
In the isolated system, this entanglement-based quantity is computed on the topological ground state of the Hamiltonian, and it results to be quantized to \(0\) in the trivial phase and \(\log2\) in the topological phase, as shown in Fig~\ref{fig:dee_gs}.
The scaling analysis confirms that the DEE is a good topological invariant in the thermodynamic limit of the system under OBCs.

Let us briefly describe what happens in a quantum-quench case. In this case, quasiparticles propagate ballistically bringing with them entanglement, and for each cut of a partition they provide an entanglement-entropy contribution given by $vt$, where $v$ is the velocity of the quasiparticles and $t$ the time~\cite{Calabrese_2005,Calabrese_2020}. So, before the propagation fronts saturate the partitions and let different cuts affect each other, one has an additional dynamical contribution to $S(A\cup B)$ of $2vt$, an additional contribution to $S(A\cap B)$ of $2vt$, an additional contribution to $S(A)$ of $vt$ and an additional contribution to $S(B)$ of $3vt$. This contributions in Eq.~\eqref{DEE} erase with each other leaving only the topological part. This cancellation starts to be disrupted after quasiparticles propagating from one cut have reached the nearest one, and the two cuts have started to affect each other so that their contributions to the entanglement are no more independent. This happens after a time $L/(4v)$ in $S(B)$, $S(A\cap B)$ and $S(A\cup B)$, being $L/4$ the minimum distance between two cuts in the partitions $B$, $A\cap B$ and $A\cup B$ (see Fig.~\ref{fig:ssh_sketch}). This is what has been observed in~\cite{Micallo_2020}, and we find a similar picture in our analytical model of Sec.~\ref{sec:kitaev_benchmark}.

In the monitored setting, the DEE is not computed on a ground state anymore. By contrast, choosing the initial state of the dynamics as the topological ground state of the Rashba nanowire model, it is computed on that state evolving in time. In particular, within the considered trajectory approach, the evolving state remains a pure state, so that the entanglement-based quantity that we study still tells us about pure quantum correlations of the system. Within the free-fermion approach discussed above, the state remains Gaussian along each quantum trajectory. Therefore, the entanglement entropy of a subsystem \(A\) can be computed from the spectrum of the reduced correlation matrix \(\mathbb G_A(t)\), obtained by restricting the full correlation matrix to the degrees of freedom belonging to \(A\) (see~\ref{appC} for details):
\begin{equation}
    \bigl[\mathbb G_A(t)\bigr]_{ij}
    =\Tr\!\left(\rho_A(t)\,\Psi_j^\dagger \Psi_i\right),
    \qquad i,j\in A\;.
\end{equation}

We evaluate the trajectory-resolved DEE, denoted by \(S^{D,\mathrm{traj}}\), for each realization of the monitored dynamics and use the resulting ensemble for the statistical analysis. In the disordered case, each trajectory is associated with an independent disorder realization, so that disorder and trajectory averages are performed simultaneously. We then compute the ensemble-averaged DEE,
\begin{equation}
    S^D=\overline{S^{D,\mathrm{traj}}}\;,
\end{equation}
as well as the probability distribution of the jump-induced variations,
\begin{equation}
    P(\Delta S^D)\;,
\end{equation}
focusing on the time window in which \(S^D\) displays a plateau.

\section{Results}
\label{sec5}
In this section we present the results of our simulations. The main goal is to characterize the time evolution of the topological edge features of the Rashba nanowire. We divide the discussion into four parts. First, we identify the spatial extent of the boundary region, namely the portion of the chain where edge effects are appreciable. We then study the time evolution of the trajectory-averaged DEE as the spatial structure of the dissipation is varied, providing a physical interpretation of the fact that the lifetime of its topological value is always linear in the system size. Then we move to the statistics of the quantum jumps and their impact on the system (Sec.~\ref{sec53}). 

\subsection{Boundary region and edge-state localization}
\label{sec51}

Before analyzing the dissipative dynamics, it is useful to determine the spatial extent of the boundary region, namely the portion of the nanowire over which edge effects are relevant. To this end, we use two complementary criteria.

\begin{figure}[ht] 
    \centering
    \includegraphics[width=.8\textwidth]{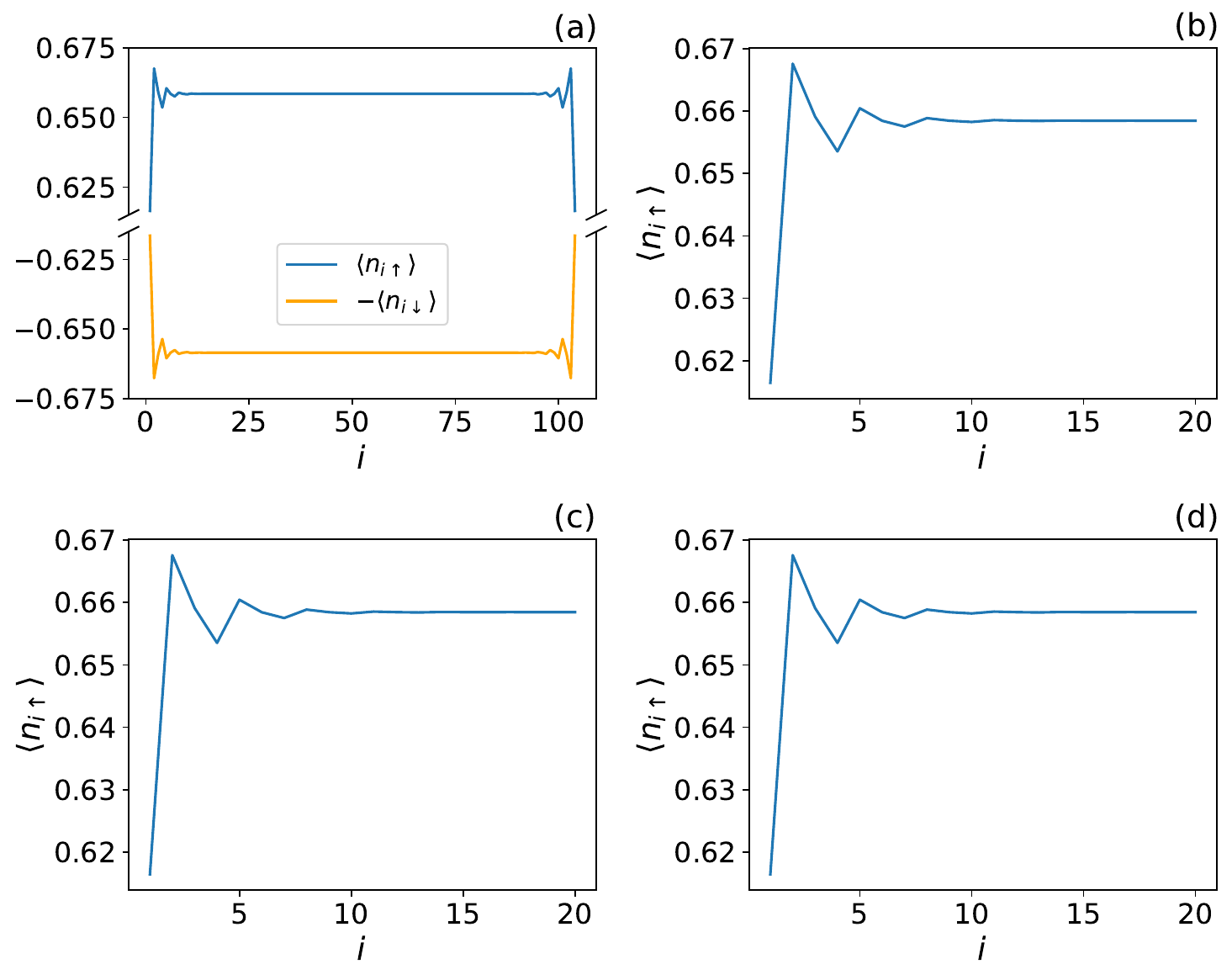}

    \caption{(a) Occupation number $\langle n_{i\sigma} \rangle$ as a function of the site index $i$, in unitary regime and with $N = 104$. (b) Occupation number $\langle n_{i\uparrow}\rangle$ of the first twenty sites in the unitary regime, with $N=88$, (c) $N=104$, (d) $N=120$.}
    \label{fig:fluttuazioni}
\end{figure}

The first criterion is based on the expectation value of a local observable in the quasiparticle vacuum, namely the local occupation number \(\langle n_{i\sigma}\rangle\). In Fig.~\ref{fig:fluttuazioni}(a), the profiles of \(\langle n_{i\uparrow}\rangle\) and \(\langle n_{i\downarrow}\rangle\) are seen to be very similar, and we therefore focus on the \(\uparrow\) component only. From the behavior of \(\langle n_{i\uparrow}\rangle\) in Figs.~\ref{fig:fluttuazioni}(b)--\ref{fig:fluttuazioni}(d), we qualitatively infer that, for system sizes \(N\in[88,120]\), boundary effects extend over approximately the first ten sites from each edge. This number does not appear to depend on the system size, at least in the considered range. 

A more direct characterization is obtained from the spatial profile of the Bogoliubov edge mode itself. Let \(\gamma_k\) denote a generic Bogoliubov quasiparticle of the Hamiltonian in Eq.~\eqref{Nanowire_rasba_Bx} (see \ref{appC} for more details), namely
\begin{equation}
    \gamma_k=\sum_{i,\sigma}\left(U_{k,i,\sigma}c_{i\sigma}+V_{k,i,\sigma}c_{i\sigma}^\dagger\right)\;.
\end{equation}
Here \(U_{k,i,\sigma}\) and \(V_{k,i,\sigma}\) represent, respectively, the particle and hole amplitudes of the \(k\)-th quasiparticle on site \(i\) and spin \(\sigma\). We then define the quasiparticle density as~\cite{10.21468/SciPostPhysLectNotes.82}
\begin{equation}
    \rho_{k,\sigma}(i)=|U_{k,i,\sigma}|^2+|V_{k,i,\sigma}|^2\;.
\end{equation}
We identify the boundary region as the portion of the chain where the zero-energy edge mode is localized. We focus on the lowest-energy Bogoliubov mode, \(k=1\), since in the topological phase it belongs to the nearly degenerate zero-energy subspace associated with the two Majorana edge modes. Its spatial density therefore provides a representative measure of the edge-state localization.

\begin{figure}[htbp]
    \centering
    \includegraphics[width=1\textwidth]{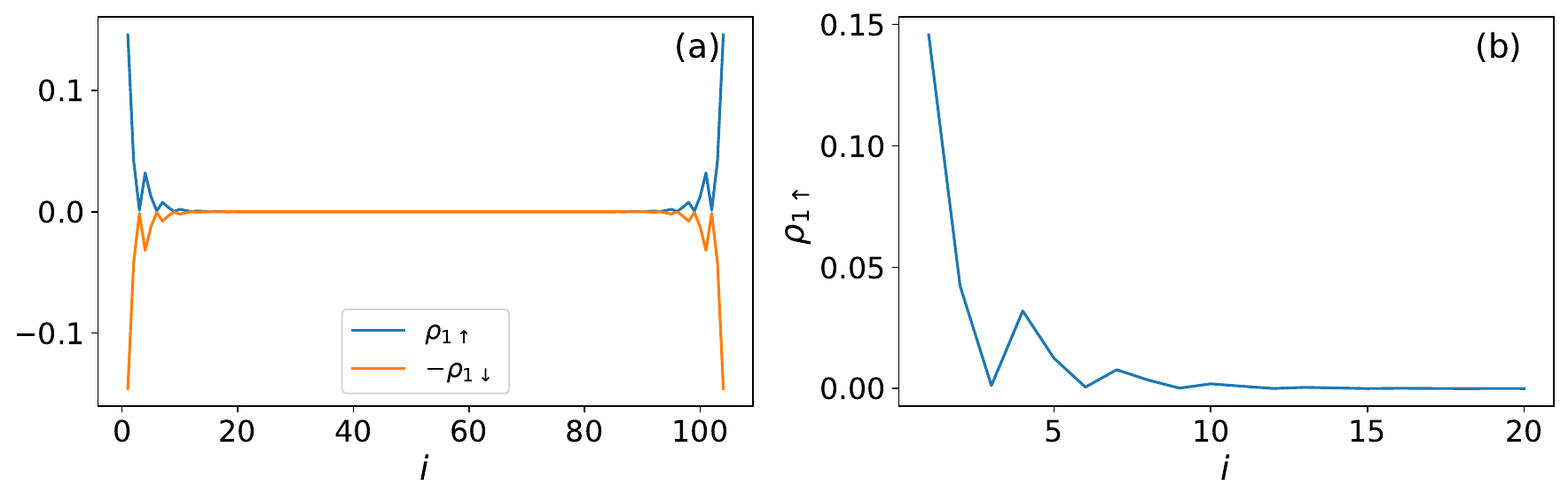}

    \caption{(a) Plot of $\rho_{1,\uparrow}$ and $\rho_{1,\downarrow}$ versus site index $i$, with $N=104$. (b) Spatial extension of $\rho_{1,\uparrow}$ for the first twenty sites.}
    \label{fig:Estensione_bordo}
\end{figure}

In Fig.~\ref{fig:Estensione_bordo}(a) we plot \(\rho_{1,\uparrow}\) and \(\rho_{1,\downarrow}\) as a function of the site index and observe that their localization profiles are essentially spin independent. We therefore restrict the analysis to \(\rho_{1,\uparrow}\). As shown in Fig.~\ref{fig:Estensione_bordo}(b), where only the first twenty sites are displayed, the quasiparticle density is localized within approximately the first ten sites. We further verified qualitatively that this localization length remains stable throughout the range \(N\in[88,120]\), corresponding to the system sizes used in our simulations.
The two criteria thus provide consistent estimates of the boundary extent. In the following, we therefore regard the first ten sites at each end of the chain as the boundary region. As explained before, in the next sections we  will consider three different monitoring protocols involving, respectively, the whole chain, the boundary region  or the bulk region (everything but the boundary). 

\subsection{Dynamics of the average DEE}\label{dydee:eqn}

We now turn to the behavior of non-local entanglement as captured by the DEE.  As explained in the Methods section, this analysis is based on a quantum-jump unraveling of the Lindblad equation, which preserves the Gaussian structure of the initial topological state. This makes it possible, following the procedure described in~\ref{appB}, to evolve the correlation matrix directly in time. The spectra of the reduced correlation matrices
\(\mathbb G_X(t)\), \(X\in\{A,B,A\cap B,A\cup B\}\), provide the corresponding
trajectory-resolved entropies \(S^{\mathrm{traj}}(X)\), from which
\(S^{D,\mathrm{traj}}\) is readily obtained.

\begin{figure}[htbp]
    \centering
    \includegraphics[width=1\textwidth]{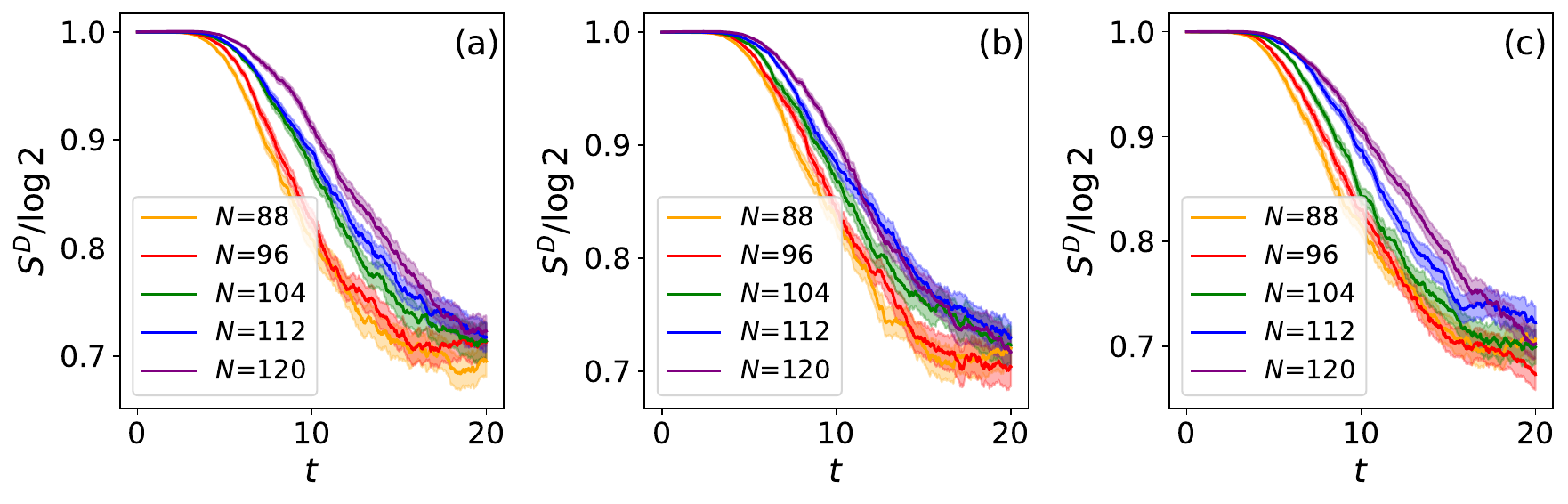}

    \caption{Evolution of the average DEE for a system prepared in the topological phase and subject to uniform dissipation. (a) $W=0$, (b) $W=0.5$, (c) $W=1.0$.}
    \label{fig:SD_unif}
\end{figure}

Averaging over an ensemble of \(N_\text{traj}=96\)
trajectories, we obtain the mean value \(S^D(t)=\overline{S^{D,\mathrm{traj}}}(t)\) together with the standard deviation of the DEE distribution. Here, the overline indicates an average over the ensemble of quantum trajectories. Figure~\ref{fig:SD_unif} shows the time evolution of the average DEE for a system initially prepared in the quasiparticle vacuum of the topological phase and subject to uniform dissipation, for different values of the disorder strength. The solid lines identify the average over the trajectories, while the shadowed areas show the standard deviations. We observe that, independently of disorder, the average DEE remains approximately constant over a time interval that increases with the system size.

\begin{figure}[htbp]
    \centering
    \includegraphics[width=1\textwidth]{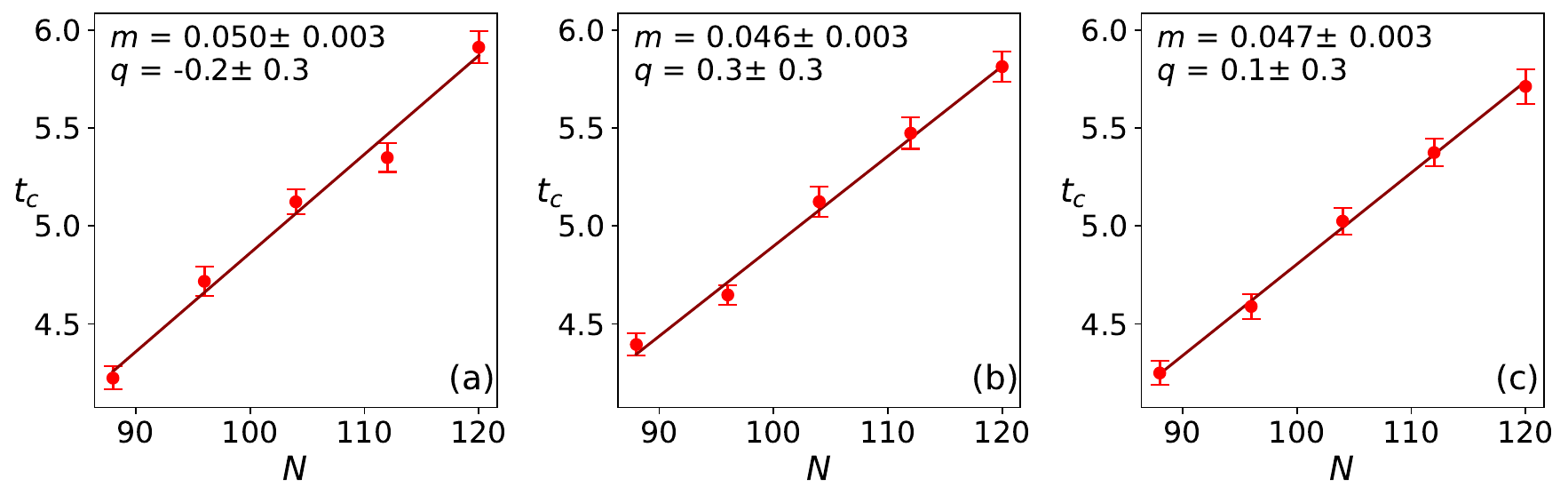}

    \caption{Characteristic time $t_c$ as a function of the system size $N$ for a system subject to uniform dissipation. The solid line represents a fit to the data, with slope $m$ and intercept $q$ indicated in each panel. (a) $W=0$, (b) $W=0.5$, (c) $W=1.0$.}
    \label{fig:tc_unif}
\end{figure}

To quantify the duration of this plateau, we introduce a characteristic time interval \([0,t_c]\) over which the average DEE remains at the topological value. More precisely, following Refs.~\cite{articolo_Giulia}, we define $t_c=\overline{t_c^{\mathrm{traj}}}$, where $t_c^{\mathrm{traj}}$ denotes, for a given trajectory, the first time at which the DEE deviates from its topological value by more than one hundredth. Accordingly, $t_c^{\mathrm{traj}}$ is determined by the threshold condition
\begin{equation}\label{sed}
\left|\overline{S^{D,\mathrm{traj}}}(t)-\overline{S^{D,\mathrm{traj}}}(0)\right|>\log 2 / 100.
\end{equation}
The results have proven to be robust under slight changes of the chosen threshold. Figure~\ref{fig:tc_unif} displays the dependence of \(t_c\) on the system size \(N\) for a system initially prepared in the quasiparticle vacuum of the topological phase and evolving under uniform dissipation. For weak disorder, the data remain compatible with a linear dependence on \(N\); we therefore fit the data with a linear law, with the resulting fit parameters shown directly in the figure. The linear increase does not depend on the specific choice of the threshold in Eq.~\eqref{sed}, as further confirmed by the fact that the curves in each panel of Fig.~\ref{fig:SD_unif} can be put on top of each other by a finite-size scaling of the horizontal axis, and the scaling factor is linear in the system size (see~\ref{resca} for details).

The linear scaling of \(t_c\) with the system size shows that the plateau of the DEE becomes longer as the chain grows, and therefore survives up to arbitrarily long times in the thermodynamic limit. This conclusion is robust against weak disorder. Moreover, ~\ref{appD} shows that restricting the dissipation either to the bulk or to the boundary does not qualitatively modify the behavior of the average DEE, both with and without disorder. 

The physical reason behind this effect -- as clarified in the analytical discussion of the Kitaev-chain toy model discussed in Sec.~\ref{sec:kitaev_benchmark} -- lies in the double degeneracy of the topological ground state and in the absence of particle conservation. Let us focus on the case where the dissipation acts only on the boundary, that is simpler and for which the difference with the SSH case is most striking. By contrast with the SSH model, where any quantum jump occurring at the boundary destroys the topological boundary mode and leads to a finite size-independent lifetime of the topological DEE, here the quantum jumps acting on the boundary let the state flip between the two degenerate topological boundary modes, alternatively creating and annihilating the topological mode. This is possible exactly because the Hamiltonian does not conserve the number of particles.

The lifetime of the topological DEE is nevertheless not infinite but scales linearly with the system size. The reason is that this flipping between the two ground state occurs together with the generation of nontopological quasiparticles near the boundaries. These quasiparticles travel ballistically along the chain and after a time linear in the system size reach the other boundary and destroy the long-range correlations between the two boundaries leading to the topological effects. Because of this quasiparticle poisoning the lifetime of the topological value of the DEE is linear in the system size, and diverges in the thermodynamic limit.

In the following subsection, we support this picture through a statistical analysis of the variation of the DEE induced by individual quantum jumps.

\subsection{Statistics of \texorpdfstring{$\Delta S^D$}{Delta SD}}
\label{sec53}

To further clarify the effect of individual quantum jumps on the DEE along a trajectory, we perform a statistical analysis of the jump-induced variations \(\Delta S^D\).

\begin{figure}[htbp]
    \centering
    \includegraphics[width=0.8\textwidth]{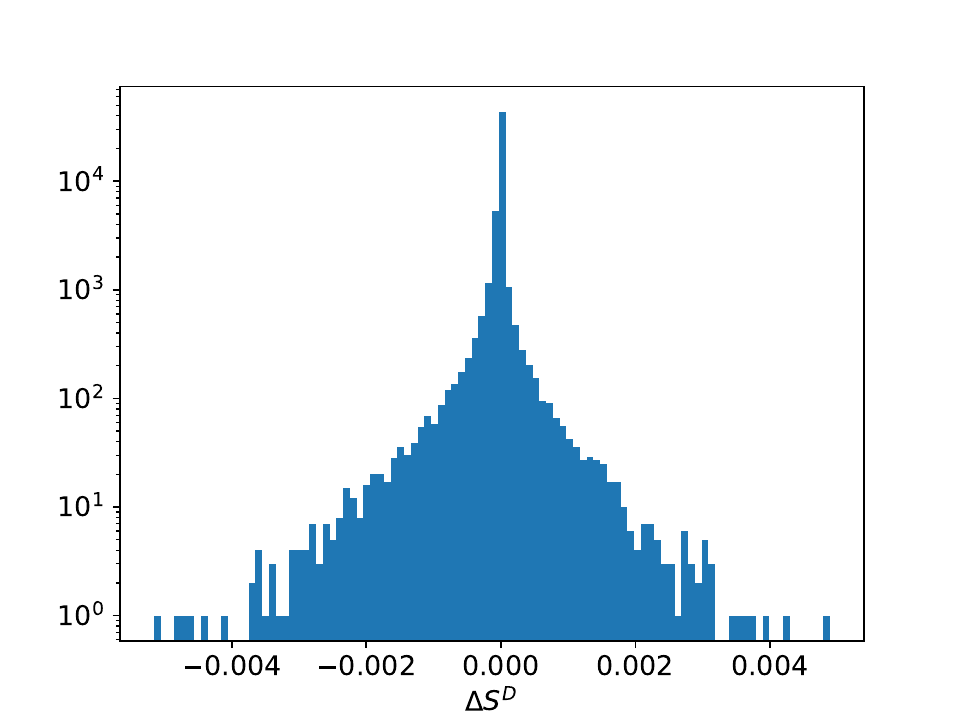}

    \caption{$P(\Delta S^D)$ obtained by evolving, for each trajectory, the system in the time interval $[0,t_c^{\text{traj}}]$ and with $N=104$.}
    \label{fig:istogrammi}
\end{figure}

Since we are interested in the transient regime in which the DEE still displays a plateau, we restrict the analysis, for each trajectory, to the time interval \([0,t_c^{\text{traj}}]\). Within this window, we record the change in the DEE after each jump event along a trajectory and repeat the procedure over many realizations in order to construct the histogram \(P(\Delta S^D)\).
In Fig.~\ref{fig:istogrammi}, we  report \(P(\Delta S^D)\) for uniform dissipation. 
In all cases, the histogram is sharply peaked at \(\Delta S^D=0\). This is fully consistent with the previous analysis, namely that the DEE does not change appreciably in the thermodynamic limit under the effect of dissipation. More precisely, these data show that quantum jumps do not instantaneously modify the value of the DEE, despite the fact that the post-jump state contains local excitations in the bulk of the chain. The non-local entanglement encoded in the disconnected partition is therefore preserved at the level of individual jump events. In the thermodynamic limit, the finite speed of propagation of quantum correlations prevents this information from reaching the entanglement cuts on finite timescales, thereby leaving the DEE unchanged during the plateau regime.
Section~\ref{sec:kitaev_benchmark} provides a paradigmatic example in the simpler Kitaev chain, where the same mechanism can be analyzed more transparently. This supports the interpretation proposed here for the Rashba nanowire, whose low-energy physics in the topological regime can be mapped onto an effective Kitaev chain~\cite{hassler2014majoranaqubits}.

\section{Kitaev-chain benchmark for the DEE under local jumps}
\label{sec:kitaev_benchmark}
In this section we use the Kitaev chain in its topological, fully dimerized limit as an analytically tractable benchmark to interpret the numerical results obtained for the dissipative Rashba nanowire. This is not meant as a microscopic derivation of the Rashba model itself. Rather, the purpose of this section is to isolate, in the simplest possible setting, the mechanism underlying the short-time robustness of the DEE under local poisoning jump events.

This viewpoint is physically motivated by the fact that the low-energy sector of the superconducting Rashba nanowire can be mapped onto an effective spinless \(p\)-wave superconducting chain, namely a Kitaev-like model~\cite{hassler2014majoranaqubits}. The Kitaev chain therefore provides a useful low-energy proxy for the edge sector relevant to the DEE, while remaining simple enough to allow for an explicit analytical treatment.

\subsection{The Kitaev chain model}

\begin{figure}
    \centering
    \includegraphics[width=\textwidth]{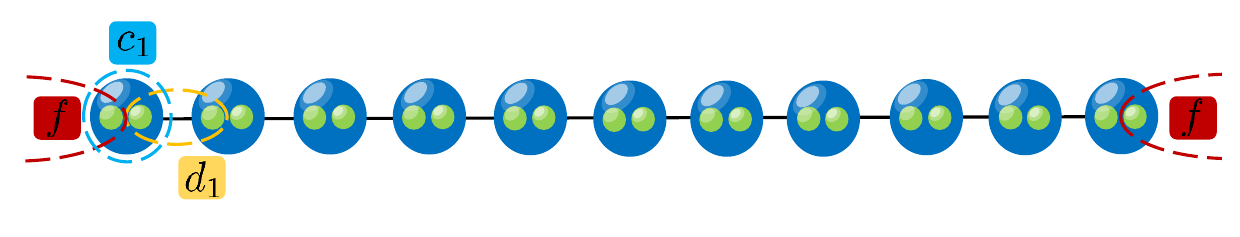}
    \caption{Sketch of the Kitaev chain in the Majorana representation. Green dots denote Majorana degrees of freedom, while blue dots denote Dirac fermions localized on the physical sites of the chain. In the trivial phase, the ground state is naturally described in terms of Dirac fermions localized on each site. In the topological phase, instead, the relevant Dirac fermions are built from Majoranas belonging to neighboring sites. In particular, the Dirac fermion localized on the physical site \(1\) is annihilated by \(c_1\), whereas the delocalized fermions are annihilated by \(d_1\) for the first bulk bond and by \(f\) for the non-local edge mode.}
    \label{fig:kitaev_sketch}
\end{figure}

We consider an open chain of $L$ spinless fermions described by the Kitaev~\cite{Kitaev_2001}
\begin{equation}
H_{\mathrm K}
={}
-\mu\sum_{j=1}^{L}
\left(c_j^\dagger c_j-\frac{1}{2}\right)
-t\sum_{j=1}^{L-1}
\left(c_j^\dagger c_{j+1}+c_{j+1}^\dagger c_j\right)
\
+
\sum_{j=1}^{L-1}
\left(
\Delta c_j c_{j+1}
+\Delta^*c_{j+1}^\dagger c_j^\dagger
\right)\;,
\label{eq:kitaev_hamiltonian}
\end{equation}
where $t$ is the hopping amplitude, $\mu$ is the chemical potential, and
$\Delta=|\Delta|e^{i\theta}$ is the superconducting pairing amplitude. The phase
$\theta$ can be removed by a global gauge transformation of the fermionic
operators, and in the following we take $\Delta$ to be real. It is convenient to resolve each physical fermion into two Majorana operators,
\begin{equation}
\gamma_{2j-1}=c_j+c_j^\dagger,
\qquad
\gamma_{2j}=i\left(c_j-c_j^\dagger\right),
\label{eq:kitaev_majoranas}
\end{equation}
which satisfy
\begin{equation}
\gamma_m^\dagger=\gamma_m,
\qquad
\left\{\gamma_m,\gamma_\ell\right\}=2\delta_{m\ell},
\end{equation}
and
\begin{equation}
c_j=\frac{1}{2}\left(\gamma_{2j-1}-i\gamma_{2j}\right).
\end{equation}
The topological phase is realized for $|\mu|<2|t|$, provided that the
superconducting gap is non-vanishing. We focus on the fully dimerized point
\begin{equation}
\mu=0,
\qquad
t=|\Delta|>0,
\label{eq:kitaev_dimerized_point}
\end{equation}
at which the Hamiltonian takes the particularly simple form
\begin{equation}
H_{\mathrm K}
=
-it\sum_{j=1}^{L-1}\gamma_{2j}\gamma_{2j+1}.
\label{eq:kitaev_majorana_hamiltonian}
\end{equation}
Thus, the Majoranas $\gamma_{2j}$ and $\gamma_{2j+1}$ are paired across the
bonds connecting neighboring physical sites, while the two Majoranas
$\gamma_1$ and $\gamma_{2L}$ remain unpaired at the ends of the chain. For every bulk bond we introduce the fermionic operator
\begin{equation}
d_j=\frac{1}{2}\left(\gamma_{2j}+i\gamma_{2j+1}\right),
\qquad
j=1,\ldots,L-1.
\label{eq:kitaev_bond_fermions}
\end{equation}
In terms of these modes, the Hamiltonian is diagonal,
\begin{equation}
H_{\mathrm K}
=
2t\sum_{j=1}^{L-1}
\left(d_j^\dagger d_j-\frac{1}{2}\right).
\label{eq:kitaev_diagonal}
\end{equation}
The modes $d_j$ are local on the bonds of the Majorana chain, although they
involve two neighboring physical sites. In particular, $d_1$ denotes the
fermion associated with the first bulk bond, formed by $\gamma_2$ and
$\gamma_3$. The two unpaired Majoranas at the physical boundaries can instead be combined
into the non-local zero-energy fermion, as depicted in Fig.~\ref{fig:kitaev_sketch}.
\begin{equation}
f=\frac{1}{2}\left(\gamma_{2L}+i\gamma_{1}\right),
\qquad
f^\dagger=\frac{1}{2}\left(\gamma_{2L}-i\gamma_{1}\right).
\label{eq:kitaev_edge_fermion}
\end{equation}
Since $f$ does not appear in Eq.~\eqref{eq:kitaev_diagonal}, its occupation
does not change the energy. The ground-state subspace is therefore
two-dimensional. We denote its two states by $\ket{0}$ and $\ket{1}$, with
\begin{equation}
\begin{gathered}
d_j\ket{0}=d_j\ket{1}=0,
\qquad
j=1,\ldots,L-1,\
f\ket{0}=0,
\qquad
\ket{1}=f^\dagger\ket{0}.
\end{gathered}
\label{eq:kitaev_ground_states}
\end{equation}
The two states differ only by the occupation of the non-local edge fermion and
belong to opposite fermion-parity sectors.

\subsection{Known results for the DEE in the topological Kitaev chain}

The basis defined by the occupation of $f$ is natural for describing the
physical boundaries of the complete chain. It is, however, not the most useful
basis for computing the reduced density matrices associated with a spatial
partition. Indeed, a bipartition introduces additional effective boundaries
at the entanglement cuts. To make this explicit, consider first a bipartition of the chain into two
connected complementary regions,
\begin{equation}
A=\left\{1,\ldots,L_A\right\},
\qquad
B=\left\{L_A+1,\ldots,L\right\}.
\end{equation}
We introduce the cut-adapted fermions
\begin{equation}
a_A=
\frac{1}{2}\left(\gamma_{2L_A}+i\gamma_1\right),
\qquad
a_B=
\frac{1}{2}\left(\gamma_{2L}+i\gamma_{2L_A+1}\right).
\label{eq:kitaev_cut_modes}
\end{equation}
The operator $a_A$ pairs the two Majoranas at the endpoints of region $A$,
whereas $a_B$ pairs those at the endpoints of its complement. We denote their
occupation numbers by
\begin{equation}
n_A=a_A^\dagger a_A,
\qquad
n_B=a_B^\dagger a_B.
\end{equation}
The construction of a cut-adapted basis for the Kitaev chain and the derivation of the DEE in the topological phase were presented in detail in Ref.~\cite{Dalmonte_PhysRevB.101.085136}, and we only recall the ingredients that are needed in the following.
The physical-edge basis and the cut-adapted basis describe the same
low-energy Hilbert space but pair the Majoranas differently. Consequently, a
state with a definite occupation of the physical-edge fermion is generally
entangled in the cut-adapted basis. With the fermionic ordering and phase
conventions adopted here, the two topological states can be written, up to
unimportant convention-dependent phases, as~\cite{Dalmonte_PhysRevB.101.085136}
\begin{subequations}
\label{eq:kitaev_cut_ground_states}
\begin{align}
\ket{0}
&=
\frac{1}{\sqrt{2}}
\left(
\ket{n_A=0,n_B=1}
-
\ket{n_A=1,n_B=0}
\right),\
\\\ket{1}
&=
\frac{1}{\sqrt{2}}
\left(
\ket{n_A=0,n_B=0}
+
\ket{n_A=1,n_B=1}
\right).
\end{align}
\end{subequations}
Tracing the first state over $B$ gives
\begin{equation}
\rho_A
=
\operatorname{Tr}_B\left[\ket{0}\bra{0}\right]
=
\frac{1}{2}
\left(
\ket{0_A}\bra{0_A}
+
\ket{1_A}\bra{1_A}
\right),
\label{eq:kitaev_reduced_density_matrix}
\end{equation}
and hence $S_A=\log 2$. The entanglement is therefore carried entirely by the
cut-induced edge degree of freedom at the fully dimerized point.

The same construction extends to the case of disconnected partitions entering the DEE.
In this case, we first have to refine the chain into consecutive connected blocks
$A_1,\ldots,A_n$ and associate with each block a cut-adapted edge fermion
$a_{A_r}$, formed from the Majoranas at its two endpoints. After a local
redefinition of the phases of the occupation states, the two topological
ground states take the general form
\begin{subequations}
    \begin{align}
        \ket{0}&=\frac{i}{\sqrt{2}}\left(-\frac{1}{\sqrt{2}}\right)^n
        \sum_{\{n_r\}_{r=1}^n}
         \left(1+(-1)^{n-\sum_r  n_{r}}\right)\ket{\{\textbf{n}\}},\\
        \ket{1}&=\frac{i}{\sqrt{2}}\left(-\frac{1}{\sqrt{2}}\right)^n
        \sum_{\{n_r\}_{r=1}^n}
         \left(1-(-1)^{n-\sum_r n_{r}}\right)\ket{\{\textbf{n}\}}.
    \end{align}
    \label{eq:gs_kitaev_partition}
\end{subequations}
Here $n_r$ is the occupation of the cut-adapted fermion associated with
$A_r$, and $n$ is the number of partitions. Hence, in the cut-adapted basis, the two topological ground states of the model can be written as equal-weight superpositions over complementary parity sectors of the edge occupations associated with the partition~\cite{Dalmonte_PhysRevB.101.085136}. As a consequence, all non-vanishing Schmidt coefficients have the same modulus, and the corresponding entanglement entropies can be read off directly. In particular, for the disconnected partition entering the definition of the DEE, one obtains the well-known topological result
\begin{equation}
    S^D=\log 2
\end{equation}
in the topological phase of the Kitaev chain. We refer to Ref.~\cite{Dalmonte_PhysRevB.101.085136} for the detailed derivation of the associated reduced density matrices. 

\subsection{Effect of a local jump on the topological ground state}
We now address the genuinely new point relevant to the present work, namely the effect of a local quantum jump on the DEE. We consider the previously-described special case of the Kitaev chain in the fully dimerized topological limit and focus only on the instantaneous effect of the jump, neglecting the non-Hermitian evolution. Although simplified, this framework captures the mechanism that underlies the transient DEE plateau observed in the dissipative Rashba nanowire.

We can rewrite the local fermionic annihilation operator at the first site as
\begin{equation}
    c_1=\frac{i}{2}\Bigl[-\bigl(d_1^\dagger+d_1\bigr)+\bigl(f^\dagger-f\bigr)\Bigr].
\end{equation}
Acting on the even-parity topological ground state \(\ket{0}\), this gives
\begin{equation}
    c_1\ket{0}
    \propto
    \,\left(-d_1^\dagger\ket{0}+\ket{1}\right)\;.
    \label{eq:postjump_state_appendix}
\end{equation}
Therefore, immediately after the jump, the system is not projected onto a generic highly excited state, but onto a superposition of two contributions: the topological ground state of opposite parity, and a state carrying one additional \emph{local} excitation on the first internal bond.

The crucial observation is that this local excitation does not immediately affect the non-local entanglement detected by the DEE. In the cut-adapted basis of Ref.~\cite{Dalmonte_PhysRevB.101.085136}, the jump only enlarges the local Hilbert-space sector associated with the block containing the excitation, while leaving unchanged the equal-weight structure of the Schmidt decomposition across the entanglement cuts. As a result, the reduced density matrices relevant for the DEE have the same non-zero spectrum as in the ground-state case, and therefore the DEE remains unchanged immediately after the jump:
\begin{equation}
    S^D_{\mathrm{post\mbox{-}jump}}=\log 2.
\end{equation}

This provides a simple analytical explanation for the numerical results of the main text. A local poisoning-like event changes the instantaneous state and creates a local excitation inside the chain, but it does not instantaneously erase the non-local edge entanglement encoded in the disconnected partition. The DEE can only change once the information carried by the local excitation has had time to propagate to the cuts defining the partition. This is precisely the mechanism behind the finite-time plateau of the DEE and its linear growth with system size observed in the dissipative Rashba nanowire.

\section{Conclusions}
\label{sec:conclusions}

We have studied the fate of topological entanglement in a monitored Rashba nanowire. Using a quantum-jump unraveling, we followed the DEE along individual pure-state trajectories and then averaged it over the trajectory ensemble. Starting from a topological ground state, we found that the topological value of the DEE survives up to a characteristic time proportional to the system size, independently of whether dissipation acts in the bulk, at the boundaries, or throughout the whole chain.

This behavior is markedly different from what happens in the SSH model. In that case, when dissipation acts at the boundaries, the topological value of the DEE survives only for a size-independent time~\cite{articolo_Giulia}. The difference can be traced back to the structure of the topological degrees of freedom. In the SSH model, the Hamiltonian conserves particle number, and the topological ground state is a Slater determinant in which the zero-energy single-particle excitation is a non-local superposition of boundary modes. This excitation is directly responsible for the topological contribution to the DEE. As a consequence, a quantum jump acting on the boundary can directly destroy it, causing the the DEE to fall to its trivial value.

In the Rashba nanowire considered here, the Bogoliubov-de Gennes Hamiltonian does not conserve particle number, and the ground state is the vacuum of quasiparticle operators, which are themselves superpositions of fermionic creation and annihilation operators.  This ground state has a degenerate topological partner, obtained by occupying the zero-energy quasiparticle mode delocalized between the two boundaries. As clarified by a simple analytical toy model based on the Kitaev chain, boundary jumps do not immediately destroy the topological contribution to the DEE. Rather, they flip the fermion parity between the two topological ground states, alternatively creating and destroying the topological mode.

However, this robustness comes at a price. Each boundary jump also creates finite-energy excitations near the edge. In the absence of bulk dissipation, these excitations move through the chain with a characteristic quasiparticle velocity. They therefore reach the regions involved in the DEE only after a time proportional to $L$. At that time, they produce extra entanglement contributions that are not topological and spoil the cancellation needed to keep the DEE quantized. As a result, the DEE moves away from its topological value. This gives a delayed quasiparticle-poisoning mechanism: unlike in the SSH case, a boundary jump does not destroy the topological contribution immediately, but only after the excitations it creates have propagated through the system. In this sense, the lifetime of the topological DEE becomes infinite only in the thermodynamic limit.

To further support this picture, we analyzed the statistics of the changes in the DEE induced by individual quantum jumps. We found no finite-value peak in the distribution of DEE variations. This is again in contrast with the SSH model, where such a peak signals the direct destruction of the topological mode and of its contribution to the DEE. 

An interesting direction for future work is to understand to what extent this mechanism is robust with respect to different unravelings of the same Lindbladian, and whether it survives beyond the free-fermion and Markovian regime. In particular, it would be important to investigate interacting systems, non-Markovian environments, and more  microscopic, experiment-driven models of poisoning processes.

\ack{
    We acknowledge useful discussions with R.~Fazio, G.~E.~Santoro and M.~M. Wauters. G.~P. and A.~R. acknowledge financial support from PNRR MUR Project PE0000023-NQSTI.}

\appendix

\section[Correlation matrix along trajectories]{Evolution of the correlation matrix along trajectories}
\label{appB}
In this appendix we provide a detailed derivation of the time evolution of the correlation function along quantum trajectories in the hypothesis that the Gaussianity of the initial topological state is preserved due to the linearity of the jumps and the quadratic non-Hermitian Hamiltonian. 

To this end, we resort to the algorithm presented in Ref.~\cite{Yip_2018} and directly apply it on the correlation matrix, retrieving the equation for $\mathbb G(t)^{\text{traj}}$. The correlation matrix that can be retrieved with the equation of motion from the Lindblad equation coincides with the average over many realizations of $\mathbb G(t)^{\text{traj}}$, namely $\overline{\mathbb G(t)^{\text{traj}}}$. In what follows, for brevity, we will denote $\mathbb G(t)^{\text{traj}}$ by $\mathbb G(t)$.

\subsection{Non-Hermitian evolution}
Let us consider a short time interval $\dt$ during which no quantum jumps occur. By recalling the stochastic Schr\"odinger equation~\eqref{equivalenza} and applying it directly to the time evolution of the correlation matrix, one finds that its deterministic non-Hermitian contribution is given by
\begin{equation}
    \begin{split}
        \mathbb G_{ij}(t+\dt)
        = \mathbb G_{ij}(t)
        + \dt \Big(
        i \braket{[H,\Psi_j^\dagger \Psi_i]}_t
        - \braket{\{\Lambda,\Psi_j^\dagger \Psi_i\}}_t
        + 2 \braket{\Lambda}_t \, \mathbb G_{ij}(t)
        \Big)
        + o(\dt)\;,
    \end{split}
    \label{eqdetG}
\end{equation}
where we introduced the shorthand
\begin{equation}
    \braket{O}_t := \bra{\psi(t)} O \ket{\psi(t)}\;,
\end{equation}
and
\begin{equation}
    \Lambda = \frac{1}{2}\sum_\mu L_\mu^\dagger L_\mu\;.
\end{equation}
To derive the non-Hermitian evolution of the correlation matrix, we write the Hamiltonian and jump operators in the Nambu-space formalism 
\begin{subequations}
    \begin{align}
    H &= \mathbf{\Psi}^\dagger \mathbb{H}\mathbf{\Psi}\;,\\
    L_\mu&=\sum_i l_{\mu i}\Psi_i\;.
    \end{align}
\end{subequations}
We then define the dissipation matrix
\begin{equation}
    M \equiv \sum_\mu L_\mu^\dagger L_\mu = \mathbf{\Psi}^\dagger \mathbb{M}\mathbf{\Psi}\;,
    \label{eq:dissipation_matrix_def}
\end{equation}
and the swap matrix
\begin{equation}
    \Sigma_x = \sigma_x\otimes 1_{2N}
    =
    \begin{pmatrix}
        0 & 1_{2N}\\
        1_{2N} & 0
    \end{pmatrix}\;.
    \label{eq:swap_matrix}
\end{equation}
The following anticommutation relations hold~\cite{10.21468/SciPostPhysLectNotes.82}
\begin{align}
    \{\Psi_i,\Psi_j^\dagger\}&=\delta_{ij}\;,\\
    \{\Psi_i,\Psi_j\}&=(\Sigma_x)_{ij}\;,\\
    \{\Psi_i^\dagger,\Psi_j^\dagger\}&=(\Sigma_x)_{ij}\;.
\end{align}.
From this definitions, we can introduce, for any matrix $\mathbb X$, the notation
\begin{equation}
    \widetilde{\mathbb X} \coloneqq \Sigma_x \mathbb X^\transpose \Sigma_x\;.
\end{equation}
Putting all together, the deterministic part of the evolution can be cast in the compact form
\begin{equation}
    \dot{\mathbb G}
    =
    i[\mathbb H,\mathbb G]
    + i\,[\widetilde{\mathbb G},\widetilde{\mathbb H}]
    + \frac{1}{2}
    \Big[
        \mathbb G (\widetilde{\mathbb M}-\mathbb M)\widetilde{\mathbb G}
        + \widetilde{\mathbb G}(\widetilde{\mathbb M}-\mathbb M)\mathbb G
    \Big]\;.
\end{equation}

\subsection{Quantum jump terms}

We now consider the contribution due to a quantum jump in the \(k\)-th channel. The probability that such a jump occurs within the time interval \(\dt\) is
\begin{equation}
    \mathrm{d}p_k
    =
    \dt \, \braket{L_k^\dagger L_k}_t
    =
    \dt \, \Tr\!\big(\mathbb M_k \mathbb G\big)\;,
\end{equation}
where
\begin{equation}
    [\mathbb M_k]_{ij} = l_{ki}^* l_{kj}\;.
\end{equation}
Accordingly, after a jump in the \(k\)-th channel, the correlation matrix is updated as
\begin{equation}
    \mathbb G(t+\dt)
    =
    \frac{1}{\dot p_k}
    \Big[
        \widetilde{\mathbb G}\,\widetilde{\mathbb M}_k\,\widetilde{\mathbb G}
        -
        \mathbb G \mathbb M_k \mathbb G
        +
        \mathbb G\,\Tr(\mathbb M_k \mathbb G)
    \Big]\;,
\end{equation}
where
\begin{equation}
    \dot p_k \coloneqq \frac{\mathrm{d}p_k}{\dt} = \Tr\big(\mathbb M_k \mathbb G\big)\;.
\end{equation}

\subsection{Evolution of the norm}

The quantum-jump unraveling algorithm requires that a jump occurs at the first time \(t^*\) such that
\begin{equation}
    \braket{\tilde{\psi}(t^*)|\tilde{\psi}(t^*)} = r\;,
\end{equation}
where \(r\) is a random number uniformly distributed in \([0,1]\), and \(\ket{\tilde{\psi}(t)}\) is the unnormalized state evolved under \(\Ham_{\eff}\). Denoting by
\begin{equation}
    n(t) \coloneqq \braket{\tilde{\psi}(t)|\tilde{\psi}(t)}\;,
\end{equation}
the squared norm of the state, its evolution over the interval \(\dt\) is
\begin{equation}
    n(t+\dt)
    =
    \braket{\tilde{\psi}(t+\dt)|\tilde{\psi}(t+\dt)}
    =
    n(t)-2\dt\,\braket{\Lambda}_t\,n(t)+o(\dt)\;.
\end{equation}
Therefore, in the limit \(\dt \to 0\), it becomes
\begin{equation}
    \frac{\mathrm{d}n(t)}{\mathrm{d}t}
    =
    -2 \braket{\Lambda}_t\, n(t)\;.
\end{equation}

\section{Properties of Gaussian states}
\label{appC}

Let us consider \(N\) fermionic modes and recall the Nambu spinor
\begin{equation}
    \mathbf\Psi=
    \big(c_1,\dots,c_N,c_1^\dagger,\dots,c_N^\dagger\big)^\transpose .
\end{equation}
A generic quadratic fermionic Hamiltonian can be written as~\cite{10.21468/SciPostPhysLectNotes.82}
\begin{equation}
    H=\,\mathbf\Psi^\dagger \mathbb H\,\mathbf\Psi
\end{equation}
where
\begin{equation}
    \mathbb H=
    \begin{pmatrix}
        A & B\\
        -B^* & -A^*
    \end{pmatrix},
    \qquad
    A^\dagger=A,\qquad B^\transpose=-B.
    \label{matriceH}
\end{equation}
The Bogoliubov--de Gennes structure in Eq.~\eqref{matriceH} implies particle--hole symmetry, so that the spectrum of \(\mathbb H\) is symmetric around zero and comes in pairs \(\{\epsilon_k,-\epsilon_k\}\). Accordingly, upon a Bogoliubov transformation
\begin{equation}
    \mathbf\Gamma=\mathbb U^\dagger \mathbf\Psi,
\end{equation}
with \(\mathbf\Gamma=(\gamma_1,\dots,\gamma_N,\gamma_1^\dagger,\dots,\gamma_N^\dagger)^\transpose\), the Hamiltonian can be diagonalized as
\begin{equation}
    H=\sum_{k=1}^N \epsilon_k \left(2\gamma_k^\dagger \gamma_k-1\right).
    \label{DiagonalH}
\end{equation}

A fermionic Gaussian state is a state fully characterized by its two-point correlation functions. Equivalently, all higher-order correlators can be reduced to products of two-point correlators by Wick's theorem. For full-rank mixed states, this is equivalent to the statement that the density matrix can be written as
\begin{equation}
    \rho=\frac{e^{-K}}{Z},
    \qquad
    Z=\Tr\big(e^{-K}\big),
    \label{rho_bella}
\end{equation}
where the entanglement Hamiltonian \(K\) is quadratic in the fermionic operators, namely \(K=\mathbf\Psi^\dagger \mathbb K \mathbf \Psi\). Pure Gaussian states are, in turn, exactly those obtained from the Fock vacuum by Gaussian unitaries, namely by evolutions generated by quadratic Hamiltonians and those that can be reached by fermionic linear optics~\cite{knill2001fermioniclinear, bravyi2004lagrangianrepresentationfermioniclinear,lumia2025phd}. Gaussianity is preserved by Bogoliubov transformations and by taking partial traces.

If \(K\) is diagonal in the Bogoliubov basis,
\begin{equation}
    K=\sum_{k=1}^N \xi_k\left(2\gamma_k^\dagger\gamma_k-1\right),
\end{equation}
then the corresponding Gaussian state factorizes as
\begin{equation}
    \rho=\bigotimes_{k=1}^N \frac{e^{-\xi_k(2\gamma_k^\dagger\gamma_k-1)}}{Z_k},
\end{equation}
with
\begin{equation}
    Z_k=\Tr\!\left[e^{-\xi_k(2\gamma_k^\dagger\gamma_k-1)}\right].
\end{equation}
In this basis, the density matrix is diagonal and each mode is occupied with probability
\begin{equation}
    \nu_k=\frac{1}{1+e^{2\xi_k}}.
\end{equation}

\subsection{Correlation matrix}

The state is therefore completely encoded in its correlation matrix
\begin{equation}
    \mathbb G_{ij}=\Tr\!\big(\rho\,\Psi_j^\dagger\Psi_i\big).
\end{equation}
In the Bogoliubov basis one has
\begin{equation}
    \mathbb G^{\scriptscriptstyle \mathrm D}
    =\mathbb U^\dagger \mathbb G\,\mathbb U
    =
    \mathrm{diag}\big(\nu_1,\dots,\nu_N,1-\nu_1,\dots,1-\nu_N\big).
    \label{Gdiag}
\end{equation}
In particular, for the quasiparticle vacuum \(\ket{0_\gamma}\), defined by \(\gamma_k\ket{0_\gamma}=0\) for all \(k\), the correlation matrix is
\begin{equation}
    \mathbb G_0=\mathbb U\,\mathbb G_0^{\scriptscriptstyle \mathrm D}\,\mathbb U^\dagger,
\end{equation}
with
\begin{equation}
    \mathbb G_0^{\scriptscriptstyle \mathrm D}
    =
    \begin{pmatrix}
        0_N & 0_N\\
        0_N & 1_N
    \end{pmatrix},
\end{equation}
where \(0_N\) and \(1_N\) denote the \(N\times N\) zero and identity matrices, respectively.

\subsection{Reduced density matrix and entanglement entropy}

Let \(A\) be a subsystem containing \(m\) fermionic modes. A key property of fermionic Gaussian states is that the reduced density matrix \(\rho_A\), obtained by tracing out the complement of \(A\), is still Gaussian. Moreover, \(\rho_A\) is completely determined by the restriction of the full correlation matrix to the modes in \(A\), which we denote by \(\mathbb G_A\).

Equivalently, there exists a quadratic entanglement Hamiltonian \(K_A\) such that
\begin{equation}
    \rho_A=\frac{e^{-K_A}}{Z_A}.
\end{equation}
If \(\{\nu_\alpha\}_{\alpha=1}^m\) are the \(m\) independent eigenvalues of \(\mathbb G_A\) (the full spectrum being \(\{\nu_\alpha,1-\nu_\alpha\}\)), then the von Neumann entanglement entropy reads
\begin{equation}
    S(A)
    =
    -\sum_{\alpha=1}^{m}
    \Big[
        \nu_\alpha \log \nu_\alpha
        +(1-\nu_\alpha)\log(1-\nu_\alpha)
    \Big].
    \label{SA_gaussian}
\end{equation}
Equivalently, in terms of the full restricted Nambu correlation matrix,
\begin{equation}
    S(A)=-\Tr\!\big(\mathbb G_A\log\mathbb G_A\big).
\end{equation}

\section{Bulk and boundary dissipations}
\label{appD}
In this appendix we show the results of the effects of dissipation confined within the bulk and at the edge of the nanowire on the DEE of the initial topological state of the model.

\begin{figure}[htbp]
    \centering
    \includegraphics[width=1\textwidth]{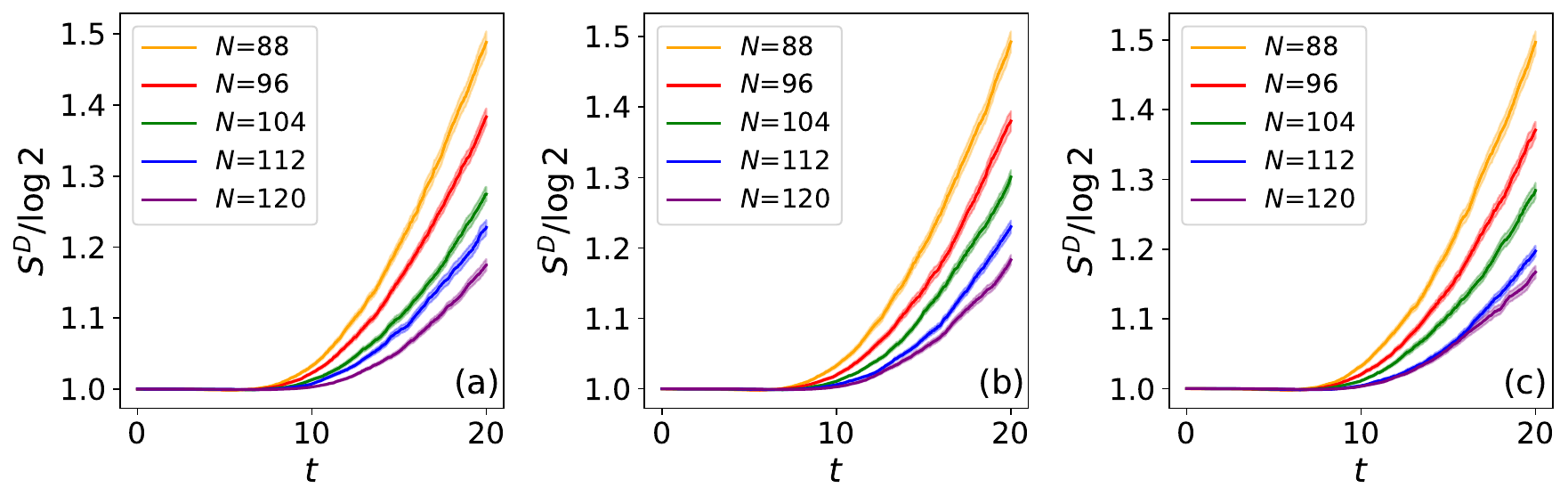}

    \caption{Evolution of the average DEE for a system prepared in the topological phase and subject to bulk dissipation. (a) $W=0$, (b) $W=0.5$, (c) $W=1.0$.}
    \label{fig:SD_bulk}
\end{figure}

Figure~\ref{fig:SD_bulk} shows the evolution of the average DEE for a system subject to bulk dissipation and initially prepared in the quasiparticle vacuum of the topological phase, for different values of the disorder strength. As in the case of uniform dissipation, we observe that the average DEE remains at the topological value over a time interval that increases with the system size, independently of disorder. 

\begin{figure}[htbp]
    \centering
    \includegraphics[width=1\textwidth]{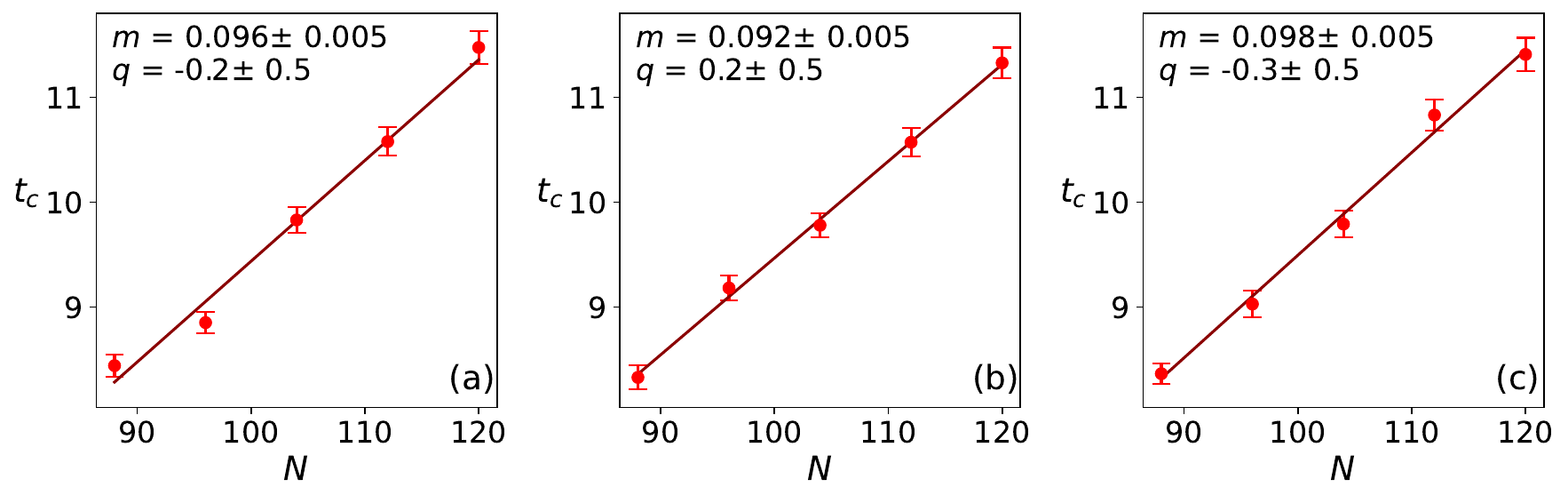}

    \caption{Characteristic time $t_c$ as a function of the system size $N$ for a system subject to bulk dissipation. The solid line represents a fit to the data, with slope $m$ and intercept $q$ indicated in each panel. (a) $W=0$, (b) $W=0.5$, (c) $W=1.0$.}
    \label{fig:tc_bulk}
\end{figure}

Figure~\ref{fig:tc_bulk} shows the linear dependence of $t_c$ on the system size $N$ for different disorder realizations. As in the case of uniform dissipation, the linear behavior is preserved in the presence of weak disorder. Linear fits have been performed for each panel, and the corresponding parameters are reported directly in the panels.

\begin{figure}[htbp]
    \centering
    \includegraphics[width=1\textwidth]{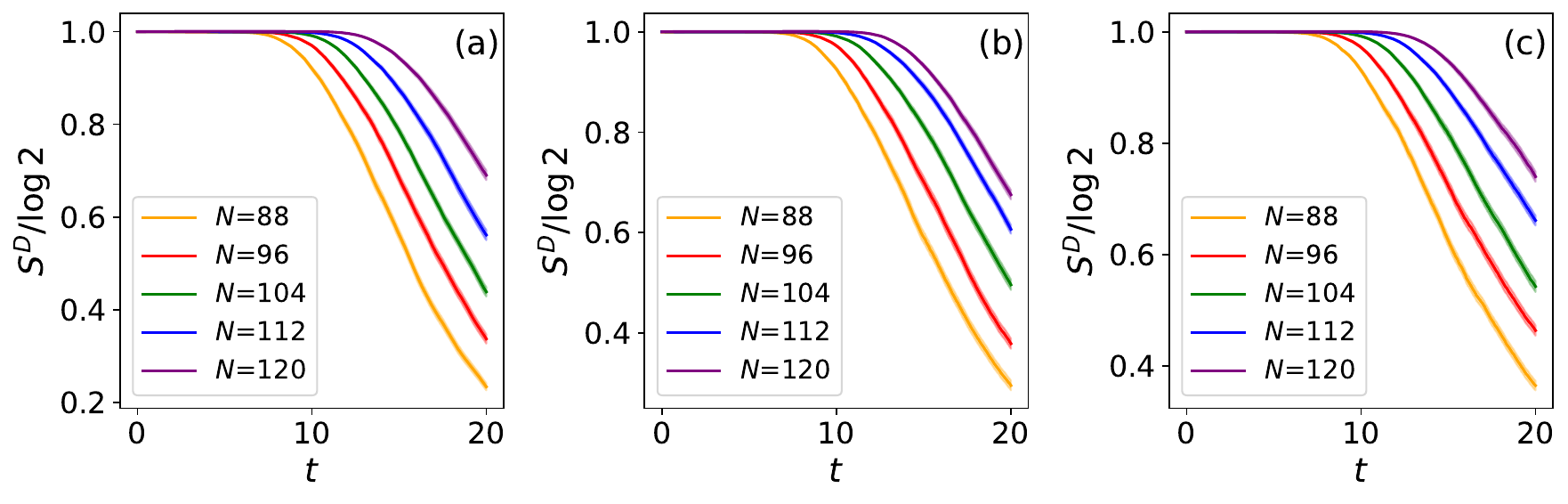}

    \caption{Evolution of the average DEE for a system prepared in the topological phase and subject to bulk dissipation. (a) $W=0$, (b) $W=0.5$, (c) $W=1.0$.}
    \label{fig:SD_bordo}
\end{figure}

Finally, in Fig.~\ref{fig:SD_bordo} we show the evolution of the average DEE for a system under boundary dissipation and initially prepared in the quasiparticle vacuum of the topological phase, for different values of the disorder strength. As in the previous cases, the average DEE remains at the topological value over a time interval that increases with the system size. 

\begin{figure}[htbp]
    \centering
    \includegraphics[width=1\textwidth]{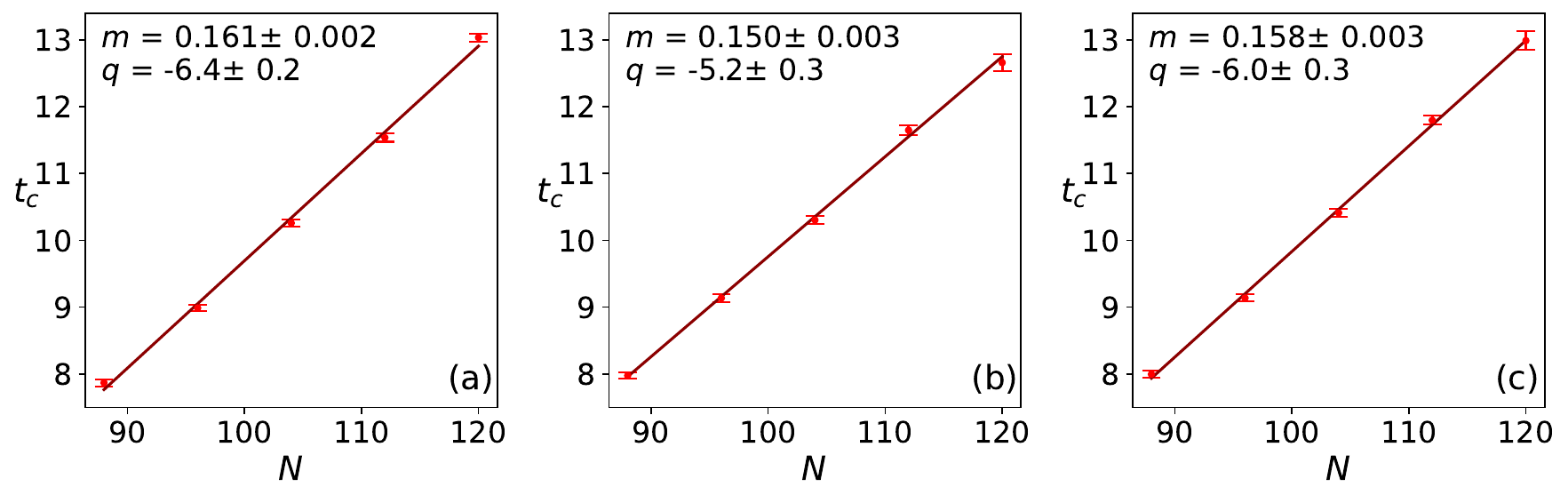}

    \caption{Characteristic time $t_c$ as a function of the system size $N$ for a system subject to boundary dissipation. The solid line represents a fit to the data, with slope $m$ and intercept $q$ indicated in each panel. (a) $W=0$, (b) $W=0.5$, (c) $W=1.0$.}
    \label{fig:tc_bordo}
\end{figure}

Figure~\ref{fig:tc_bordo} shows the linear dependence of $t_c$ on the system size $N$ for different realizations of disorder. We thus observe, as in the previous cases, that the linear trend is preserved in the presence of weak disorder. The parameters of these relations, obtained from linear fits, are are indicated in the panels.

\begin{figure}[htbp]
    \centering
    \includegraphics[width=1\textwidth]{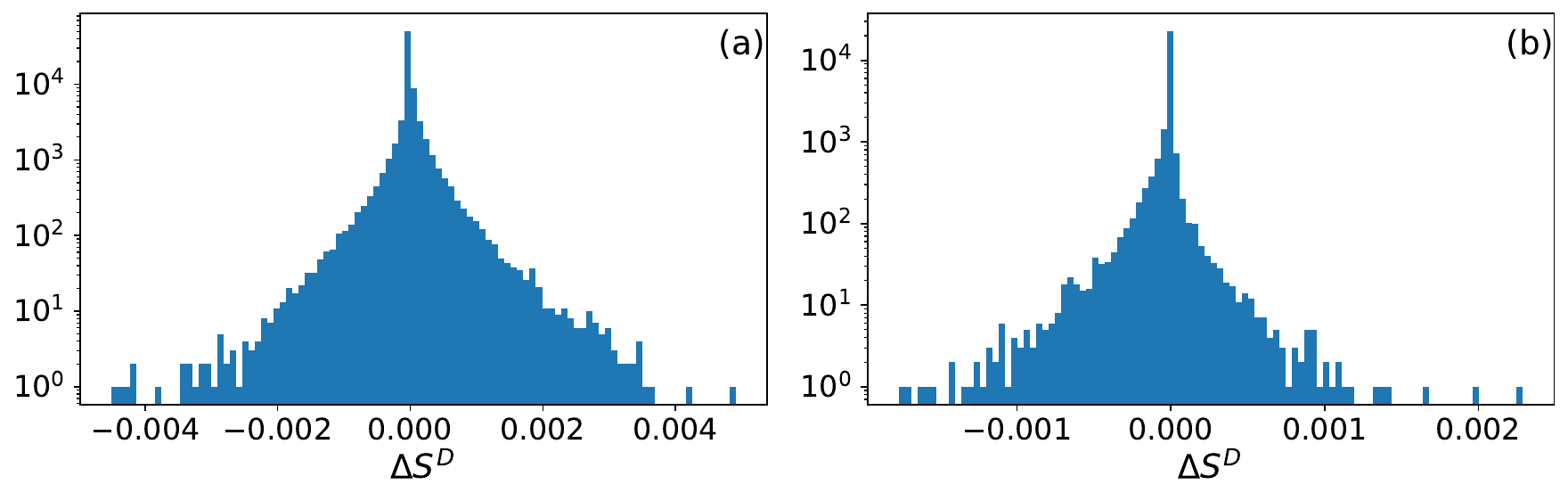}

    \caption{$P(\Delta S^D)$ obtained by evolving, for each trajectory, the system in the time interval $[0,t_c^{\text{traj}}]$ and with $N=104$. (a) Bulk dissipation, (b) Boundary dissipation.}
    \label{fig:istogrammi_boundary_bulk}
\end{figure}

Finally, in Fig.~\ref{fig:istogrammi_boundary_bulk}, we report the probability distribution $P(\Delta S^D)$ for both bulk and boundary dissipation. As in the case of uniform dissipation, the histogram is sharply peaked around $\Delta S^D = 0$. This indicates that, even when the spatial profile of the dissipation is modified, the statistical properties of $\Delta S^D$ and the dynamical evolution of the DEE lead to the same physical picture.

\section{Time rescaling}
\label{resca}

\begin{figure}[h!]
    \centering
    \includegraphics[width=\textwidth]{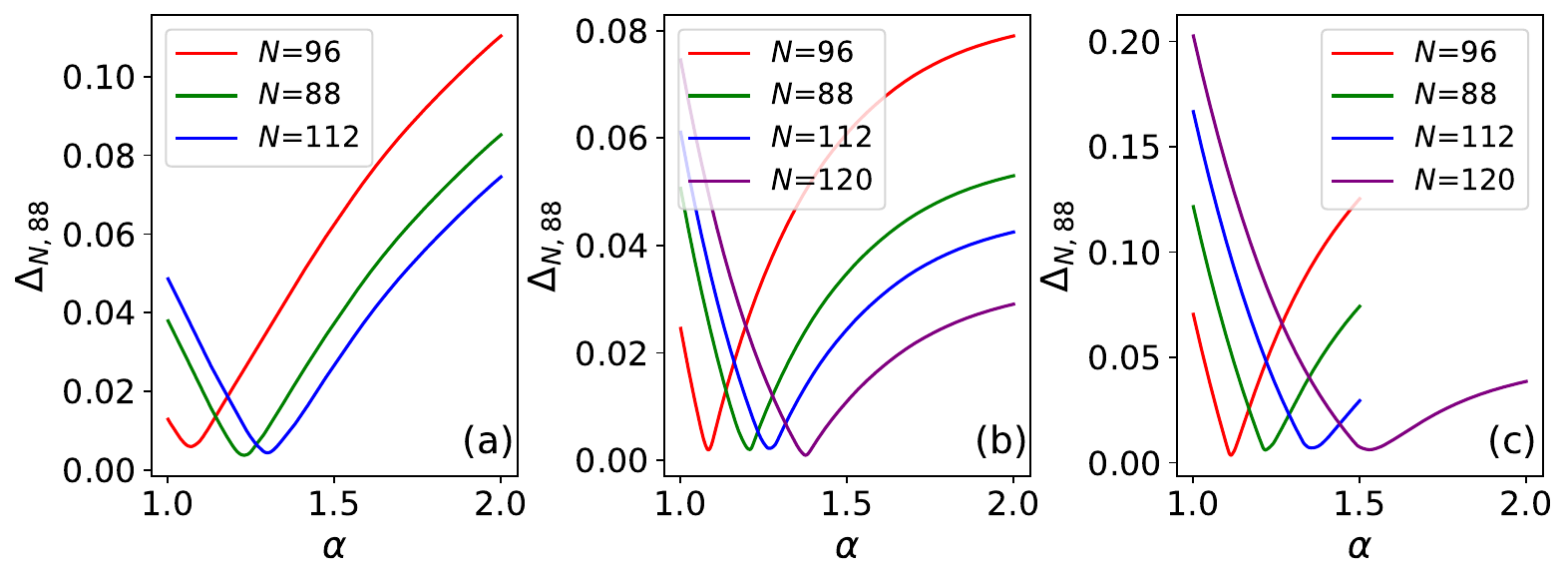}

    \caption{Distance between the rescaled curve at system size \(N\) and the corresponding reference curve (\(N=88\)), for different values of $N$ in the cases of uniform dissipation (a), bulk dissipation (b), and boundary dissipation (c).}\label{minimum:fig}
\end{figure}

\begin{figure}[h!]
    \centering
    \includegraphics[width=\textwidth]{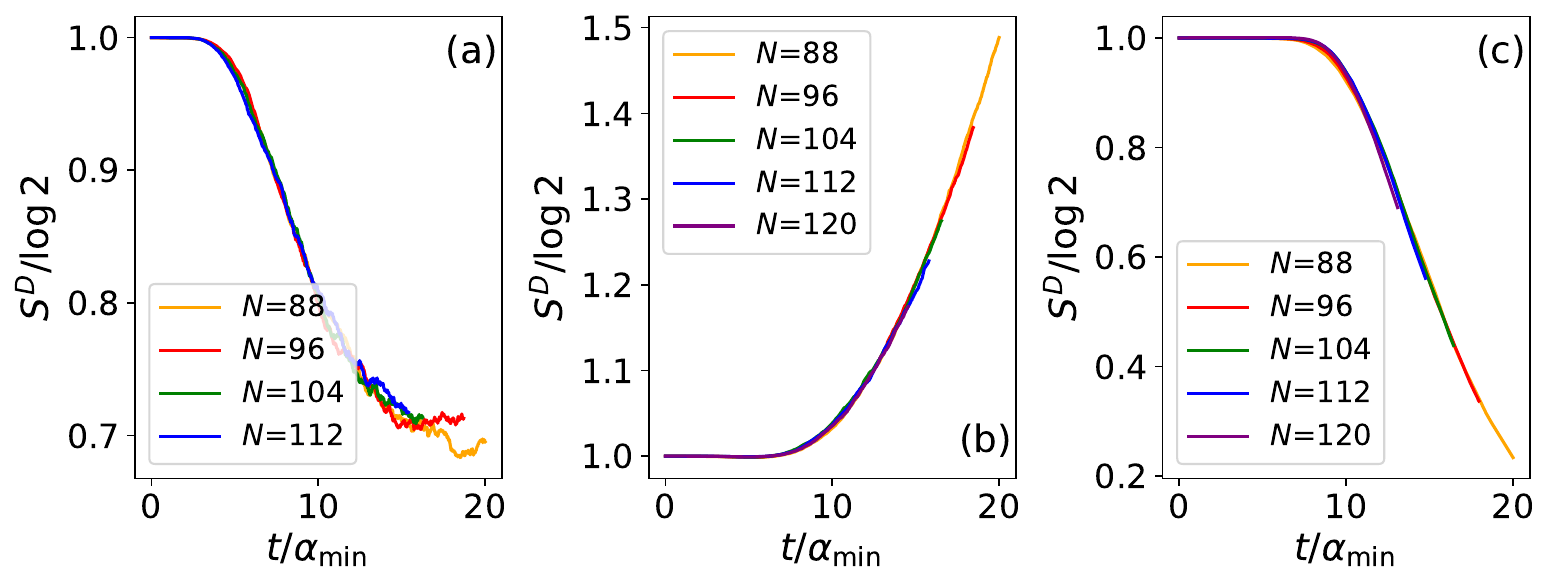}

    \caption{$S^D$ versus the rescaled time $t/\alpha_{\text{min}}$ in the cases of uniform dissipation (a), bulk dissipation (b), and boundary dissipation (c). (For $N=88$ the rescaling factor is 1.) }\label{imba_resc:fig} 
\end{figure}

\begin{figure}[h!]
  \begin{center}
    \includegraphics[width=80mm]{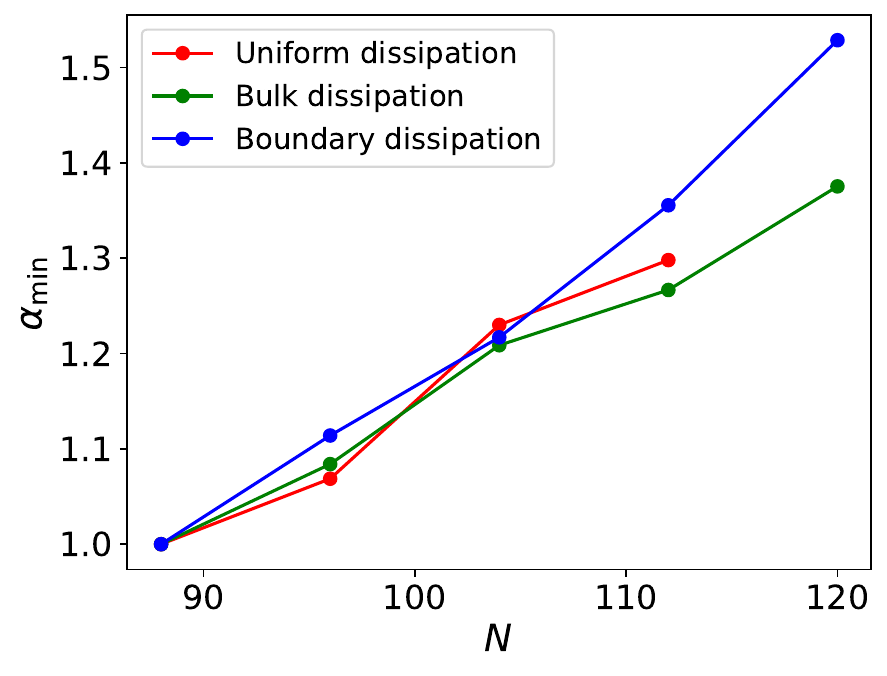}
    \caption{$\alpha_{\rm min}$ as a function of $N$ al in the three chosen dissipation extensions.}\label{alphamin:fig}
  \end{center}
\end{figure}

The DEE time traces for different system sizes (see Fig.~\ref{fig:SD_unif}) seem to be the same curve with time rescaled by a different factor. In order to quantitatively show this fact we perform a rescaling. Specifically, we rescale the time $t$ by a factor $\alpha$ such that the rescaled DEE is $S^D_{r,\alpha}(N,t/\alpha) \equiv S^D(N,t)$. We evaluate the distance between the rescaled curve at size $N_1$ and the reference curve at size $N_2$
\begin{align}
  \Delta_{N_1,N_2} &= \frac{\int |S^D_{r,\alpha}(N_1,t')-S^D(N_2,t')|\dd t'}{\sqrt{\int |S^D_{r,\alpha}(N_1,t')|\dd t'\int |S^D(N_2,t')|\dd t'}} \nonumber\\
  &=\frac{\int |S^D(N_1,\alpha t')-S^D(N_2,t')|\dd t'}{\sqrt{\int |S^D(N_1,\alpha t')|\dd t'\int |S^D(N_2,t')|\dd t'}} \,,
\end{align}
and we minimize it. The minimum is very clear, in all the considered cases, as we can see Fig.~\ref{minimum:fig}, where we plot $\Delta_{N,88}$ versus $\alpha$ for different values of $N$ and the three considered cases. Let us call this minimum $\alpha_{\rm min}$. In Fig.~\ref{imba_resc:fig} we plot the curves $S^D(N,t)$ versus the rescaled time $t/\alpha_{\rm min}$ and we see that the agreement between the different curves is pretty good. In Fig.~\ref{alphamin:fig} we plot $\alpha_{\rm min}$ versus $N$ and we see that it increases more or less linearly, but larger sizes are needed in order to give a ultimate statement. In summary, these results show that the decay time -- that is proportional to $\alpha_{\rm min}$ due to the overlap of the rescaled curves -- linearly increases with $N$, for the system sizes we have access to. 

    \clearpage
    \phantomsection
    \bibliographystyle{iopart-num}
    \bibliography{apssamp}

\end{document}